\documentclass[reprint,superscriptaddress,preprintnumbers,amsmath,amssymb,aps,prd,nofootinbib]{revtex4-2}

\usepackage{placeins} 
\usepackage[normalem]{ulem} 
\usepackage{graphicx}  
\usepackage{xcolor}
\usepackage{dcolumn}   
\usepackage{bm}        
\usepackage{hyperref}  
\usepackage{multirow}
\usepackage{siunitx}
\graphicspath{
  {./figures/}
}

\def\ga{\gamma}

\def\ze{\zeta}

\def\la{\lambda}

\def\si{\sigma}

\def\De{\Delta}

\def\La{\Lambda}

\def\half{\frac{1}{2}}

\def\bx{{\mathbf{x}}}
\def\by{{\mathbf{y}}}
\def\bz{{\mathbf{z}}}

\def\bX{{\mathbf{X}}}

\newcommand{\ben}{\begin{equation}}
\newcommand{\een}{\end{equation}}
\newcommand{\bea}{\begin{eqnarray}}
\newcommand{\eea}{\end{eqnarray}}
\newcommand{\ba}{\begin{array}}
\newcommand{\ea}{\end{array}}
\newcommand{\bit}{\begin{itemize}}
\newcommand{\eit}{\end{itemize}}

\newcommand{\IniCorLen}{l_{\phi}}

\newcommand{\Dx}{\Delta x}

\newcommand{\FitCon}{k} 
\newcommand{\FitSlo}{m} 

\newcommand{\Lwind}{\ell_\text{w}}

\newcommand{\ms}{m_\text{s}} 

\newcommand{\SlenRes}{\ell_\text{r}} 

\newcommand{\tStart}{\ensuremath{\tau_{\rm start}}}
\newcommand{\tEnd}{\ensuremath{\tau_{\rm end}}}
\newcommand{\tHalf}{\ensuremath{t_{L/2}}}
\newcommand{\tDiff}{\ensuremath{\tau_{\rm diff}}}
\newcommand{\tcg}{\ensuremath{\tau_{\rm cg}}}

\newcommand{\Tc}{T_\text{c}}

\newcommand{\vol}{\mathcal{V}}
\newcommand{\vAv}{v}

\newcommand{\vs}{v_\text{s}}
\newcommand{\vLAsymEst}{\hat{v}_{L,*}}

\newcommand\vev[1]{\left\langle #1 \right\rangle}

\newcommand{\WtFun}{\mathcal{W}}

\newcommand{\ws}{w_\text{s}}

\newcommand{\xr}{x_{\rm r}}

\newcommand{\xir}{\xi_{\rm r}}

\newcommand{\xiw}{\xi_{\rm w}}

\newcommand{\xLinEst}{\hat{x}_\text{lin}}
\newcommand{\xrLinEst}{\hat{x}_\text{r,lin}}
\newcommand{\xwLinEst}{\hat{x}_\text{w,lin}}
\newcommand{\xrAsymEst}{\hat{x}_{\text{r},*}}

\newcommand{\zer}{\ze_{\rm r}}
\newcommand{\zew}{\ze_{\rm w}}

\newcommand{\zerAsymEst}{\hat{\ze}_{\text{r},*}}
\newcommand{\zewAsymEst}{\hat{\ze}_{\text{w},*}}
\newcommand{\zerLinEst}{\hat{\ze}_\text{r,0}}
\newcommand{\zewLinEst}{\hat{\ze}_\text{w,0}}

\newcommand{\xStarRestVOS}{0.814}
\newcommand{\dxStarRestVOS}{37}
\newcommand{\vsStarVOS}{0.609}
\newcommand{\dvsStarVOS}{14}

\newcommand{\zzRestMeansVOS}{1.50}
\newcommand{\zzRestErrsVOS}{11}
\newcommand{\zzWindMeansVOS}{1.20}
\newcommand{\zzWindErrsVOS}{9}

\newcommand{\zzWindMeanAxCESS}{1.220}
\newcommand{\zzWindErrsAxCESS}{57}
\newcommand{\zzRestMeanAxCESS}{1.491}
\newcommand{\zzRestErrsAxCESS}{93}
\newcommand{\xRestMeanAxCESS}{0.822}
\newcommand{\xRestErrsAxCESS}{19}
\newcommand{\vLMeanAxCESS}{0.5705}
\newcommand{\vLErrsAxCESS}{93}

\definecolor{darkgreen}{rgb}{0.0,0.7,0.0}

\begin{document}

\preprint{HIP-2024-22/TH}

\title{Scaling density of axion strings in terasite simulations}

\newcommand{\Sussex}{\affiliation{
Department of Physics and Astronomy,
University of Sussex, Falmer, Brighton BN1 9QH,
U.K.}}

\newcommand{\HIPetc}{\affiliation{
Department of Physics and Helsinki Institute of Physics,
PL 64, 
FI-00014 University of Helsinki,
Finland
}}

\newcommand{\EHU}{\affiliation{
Department of Physics,
University of the Basque Country UPV/EHU, 
48080 Bilbao,
Spain
}}

\newcommand{\addressEHUQC}{\affiliation{EHU Quantum Center, University of the Basque Country UPV/EHU, Leioa, 48940 Biscay, Spain}}

\newcommand{\Tufts}{\affiliation{
Institute of Cosmology, Department of Physics and Astronomy, 
Tufts University,
Medford, MA 02155,
USA}}

\author{Jos\'e Correia}
\email{jose.correia@helsinki.fi}
\HIPetc

\author{Mark Hindmarsh}
\email{mark.hindmarsh@helsinki.fi}
\HIPetc
\Sussex

\author{Joanes Lizarraga}
\email{joanes.lizarraga@ehu.eus}
\EHU
\addressEHUQC

\author{Asier~Lopez-Eiguren}
\email{asier.lopez@ehu.eus}
\EHU
\addressEHUQC

\author{Kari Rummukainen}
\email{kari.rummukainen@helsinki.fi}
\HIPetc

\author{Jon Urrestilla}
\email{jon.urrestilla@ehu.eus}
\EHU
\addressEHUQC

\date{\today}

\begin{abstract}
We report on a study of axion string networks using fixed-grid simulations of up to $16384$ points per side. The length of string can be characterised in terms of standard dimensionless parameters $\zeta_\text{w}$ and $\zeta_\text{r}$, the length density measured in the cosmic rest frame and the string rest frame, scaled with the cosmic time $t$. The motion of the string can be characterised by the root-mean-square (RMS) velocity of the string. Starting from a range of initial length densities and velocities, we analyse the string network in the standard scaling framework and find evolution towards a fixed point with estimated values $\hat{\zeta}_{\text{w},*} =  1.220(57)$ and $\hat{\zeta}_{\text{r},*} = 1.491(93)$. The two measures are related by the RMS velocity, which we estimate to be $\hat{v}_{*} = 0.5705(93)$. The length density is consistent with previous measurements, while the velocity is about 5\% lower. For simulations starting from low enough density, the length density parameters $\zeta_\text{w}$ and $\zeta_\text{r}$ remain below their fixed point values throughout, while growing slowly, giving rise to an impression of approximately logarithmic increase with time. This has been proposed as the true long-term behaviour. We find that the growth tends to slow down as the values of $\zeta_\text{w}$ and $\zeta_\text{r}$ identified as fixed points are approached.  In the case of $\zeta_\text{r}$, the growth stops for simulations which started close to the fixed point length density. The difference between  $\zeta_\text{w}$ and $\zeta_\text{r}$ can be understood to result from the continuing velocity evolution. Our results indicate that the growth of $\zeta_\text{w}$ is a transient appearing at low densities and while the velocity is converging. This highlights the importance of studying the string density and the velocity together, and the preparation of initial conditions.
\end{abstract}

\maketitle

\section{Introduction}

The axion, a very weakly coupled pseudoscalar particle, emerges from the Peccei-Quinn (PQ) mechanism  \cite{Weinberg:1977ma,Wilczek:1977pj,Peccei:1977hh,Peccei:1977ur} which solves the strong CP problem of Quantum Chromodynamics (QCD). The PQ mechanism proposes a new global anomalous abelian symmetry U(1)$_{\rm PQ}$ which is spontaneously broken at an energy scale well above that of the Standard Model. The higher the scale, the more weakly the axion is coupled to Standard Model particles,  \cite{Kim:1979if,Shifman:1979if,Zhitnitsky:1980tq,Dine:1981rt}, and the axion becomes a promising dark matter candidate  \cite{Preskill:1982cy,Abbott:1982af,Dine:1982ah} for 
PQ symmetry-breaking in the range $10^{10}$ GeV to $10^{12}$ GeV (see e.g.~\cite{Marsh:2015xka,Sikivie:2020zpn,Irastorza:2021tdu} for reviews). 
If PQ symmetry breaking happens after primordial inflation, then 
axion cosmic strings \cite{Vilenkin:1982ks,Davis:1986xc} are produced. Axion strings are a kind of global cosmic string  \cite{Hindmarsh:1994re,Vilenkin:2000jqa}, 
pairs of which become tied together by domain walls when the temperature of the universe falls to the QCD scale, around 100 MeV.
The majority of the energy of the axion string network is left behind in the form of axion waves. In order to calculate the dark matter density, these waves must be accurately characterised in amplitude and frequency. 

The dynamics of a network of axion strings, and the axion radiation that they produce, follow a set of non-linear field equations coming from a scalar field theory in the classical approximation, which are best studied using numerical simulations. 
Several different groups have been tackling this endeavour
\cite{Yamaguchi:1998gx,Yamaguchi:1999wt,Yamaguchi:2000fg,Yamaguchi:2002sh,
Hiramatsu:2010yu,Hiramatsu:2012gg,Kawasaki:2014sqa,Fleury:2015aca,Lopez-Eiguren:2017dmc,Klaer:2017qhr,Klaer:2017ond,Gorghetto:2018myk,Kawasaki:2018bzv,Vaquero:2018tib,Buschmann:2019icd,Klaer:2019fxc,Hindmarsh:2019csc,Gorghetto:2020qws,Gorghetto:2021fsn,Buschmann:2021sdq,Benabou:2023ghl,Saikawa:2024bta,Kim:2024wku}. While all simulations give compatible results there is still disagreement about interpreting them and extracting the late time behaviour of the network \cite{Hindmarsh:2021zkt}. The disagreement has significant impact on calculations of the axion mass in the case where it constitutes all the dark matter. Precise estimates are needed for experimental searches, which rely on resonant detectors.

Here we extend our previous work on the numerical study of axion string networks, concentrating on the key property of scaling. 
When a string network is scaling, all length scales describing the network in a statistical sense are proportional to the Hubble length, and of order the Hubble length, apart from the string width which remains microscopic. 
Scaling follows from the hypothesis that the string tension can be factored out of the dynamics of the string network as its radius of curvature grows much larger than the string width. Dimensionless ratios of length scales to the Hubble length are therefore constant, and one can extrapolate network quantities to arbitrarily late times if one can measure the ratios. 

One quantity that can be used to characterise scaling is the dimensionless length density parameter of axion string $\zeta$,
which is one quarter of the mean number of horizon lengths of string per horizon volume. 
A constant value of the parameter at late times is a prediction of scaling.
All simulations have $\zeta \simeq 1$ by the end, from a wide variety of initial length densities, which is consistent with scaling. The value of $\ze$ at the end of the simulation is generally approached from below. 

Other groups have argued that a correction to scaling is needed to explain the numerical simulations \cite{Gorghetto:2018myk,Kawasaki:2018bzv,Vaquero:2018tib,Buschmann:2019icd,Klaer:2019fxc,Gorghetto:2020qws,Buschmann:2021sdq,Saikawa:2024bta,Kim:2024wku}
and that $\zeta$ will continue to grow, in proportion to the logarithm of cosmic time \cite{Gorghetto:2018myk,Gorghetto:2020qws}, or a more complicated function \cite{Klaer:2019fxc,Saikawa:2024bta}, 
or asymptote to a higher value \cite{Martins:2018dqg}. 
If this were to be true, the density of string would be larger at the QCD scale than numerical simulations indicate, and, in the case of continued growth, much larger, as the natural logarithm of the ratio of the PQ scale to the QCD scale is around 70.  Estimates of the string density at the QCD scale depend on the extrapolation method, but groups advocating logarithmic growth estimate that it will be greater by a factor of about 10 \cite{Gorghetto:2020qws,Buschmann:2021sdq}. The resulting axion number density seems not to be very sensitive to the string density at the QCD scale \cite{Dine:2020pds,Klaer:2017ond}, but axion dark matter mass estimates in the various long-term growth scenarios are still uncertain, ranging from $26$ $\mu$eV \cite{Klaer:2017ond} to as much as 450 $\mu$eV \cite{Saikawa:2024bta}.

The only significant difference between the approaches of different groups is the preparation of the initial conditions of the fields, which determines the initial string density. 
When the initial length density parameter $\zeta$  is much lower than 1, it increases significantly throughout the simulation. 
The approach of $\zeta$ towards unity from low densities was highlighted in Ref.~\cite{Gorghetto:2018myk}, fitted to logarithmic growth, and interpreted as the long-term behaviour of the network.  The approximately logarithmic growth from low initial string densities was subsequently observed by all groups.

We also observed slow growth from low density, and also that the final $\zeta$ depends on the initial string density.  We argued that this is to be expected in finite volume simulations, where there is not enough time to reach the true scaling density, and showed that the scaling value of $\zeta$ can be extracted with $\mathcal{O}(10\%)$ accuracy \cite{Hindmarsh:2019csc,Hindmarsh:2021vih,Hindmarsh:2021zkt}. 

In studying the system, it is important to distinguish between the length density parameter measured in the local rest frame of the string $\zer$ and in the cosmological rest frame (or universe frame) $\zew$. 
The latter is Lorentz contracted, and therefore affected by the evolution of the RMS velocity of the string network $v$. We will see that $v$ also slowly converges to a fixed point, and when the evolution is from higher RMS velocities, the Lorentz contraction decreases throughout the simulation and exaggerates the time-dependence of  $\zew$. 

There is a clear-cut test to check whether standard scaling or logarithmic growth is the true evolution. In the case of logarithmic growth, the value of $\zeta$ will continue to grow, and thus get values which significantly exceed our estimate of the asymptotic length density parameter $\zewAsymEst= \zzWindMeansVOS(\zzRestErrsVOS)$. 
No clear convergence in $\zeta$ has been observed, but neither has the universe-frame asymptotic length density parameter been exceeded. 

One way to resolve the issue is by having a larger dynamical range.
The largest fixed-grid simulations to date are those of Ref.~\cite{Saikawa:2024bta}, carried out on a lattice of size $11264^3$, and those of Ref.~\cite{Buschmann:2021sdq}, carried out on an adaptive mesh. Both sets of simulations reached 
a horizon-to-string width ratio of around $8000$, and in both $\zeta \simeq 1.2$ (in the cosmic rest frame) at the end of the simulations. These simulations therefore do not contradict the predictions of the standard scaling model, although convergence in the string length density was not manifest.

We have performed radiation-era cosmological simulations in fixed cubic grids of up to $16384$ (16k) sites per side. 
This has been possible thanks to HILA \cite{HILA}, a simulation framework designed for very large-scale simulations on a variety of supercomputing architectures, including GPGPU machines. 

With this increased dynamic range we are able to demonstrate the explicit convergence of $\zer$, 
by which we mean $\zer$ statistically constant in time over the last half of the simulation, in simulations which start with densities $\zer \gtrsim 1.5$, close to or above the final value. 
Moreover, the final values are consistent with our previous simulations on smaller lattices, which used linear fits to the mean string separation against time to estimate the asymptotic length density.  
The linear fit method also gives values of the asymptotic $\zer$ agreeing with its value at half the light-crossing time of the simulation box.

We see that the initial values of $\ze$ and $v$ are important in reaching explicit convergence by the end of a simulation, as a finite-volume simulation is also a finite-duration simulation. As mentioned above, 
explicit convergence is observed in the rest-frame length density in simulations starting at higher density. Conversely, the velocity estimator convergence is observed in the lowest density simulations, starting at $\zer = 0.16$.   RMS velocities measured for higher initial string densities evolve throughout the simulations, but in the case of the largest simulations they agree by the end. 
They are about 5\% lower than our previous estimates on smaller lattices, and outside the 1$\si$ uncertainty range, indicating that uncertainties in the velocity were underestimated.

We also look for behaviour consistent with long-term growth. 
At values of $\zew$ around $0.5$, we find that $\zew$ grows close to proportionally with the logarithm of time, and the coefficient is consistent with other groups' results ($\zer$ is not commonly measured). However, as $\zew$ grows beyond $1$, the slopes tend to decrease.  Furthermore, the slope of $\zer$ against the logarithm of time decreases quite strongly, consistently with an approach to a fixed point. The difference between the two length measures is a result of the velocity evolution changing the Lorentz contraction of the cosmic rest frame measure.

In summary, our results support the standard scaling picture, and give a more precise measurement of the scaling density and the scaling RMS velocity.  The convergence of the rest-frame length density parameter for simulations which start near the scaling density is strong evidence in support of standard scaling.  Our results also highlight the importance of initial conditions, and suggest that even clearer convergence could be obtained from simulations which start close to the scaling velocity as well. 

Scripts and data needed to generate all figures and tables derived from the simulations can be found at the following location \cite{correia_2024_14413516}.


\section{Model and observables}
\label{sec:ModelSims}

The simplest axion models \cite{Kim:1979if,Shifman:1979if,Zhitnitsky:1980tq,Dine:1981rt} 
add to the Standard Model a singlet scalar field with a U(1) symmetry, $\Phi$, with action
\ben
S=\int d^4 x \sqrt{-g} \Big( \frac{1}{2} \partial_{\mu} \Phi^T \partial^{\mu}\Phi- \frac{1}{4}\lambda(\Phi^T\Phi-\eta^2)^2 \Big),
\label{eq:ac}
\een
where we have written the field as a two-component vector, and the U(1) symmetry is realised as a rotation on the vector. 
At a temperature $\Tc \simeq \eta$, the model undergoes a phase transition, spontaneously breaking the U(1) symmetry.
The magnitude of the field takes the equilibrium value
$\eta$, with a massless pseudoscalar fluctuation mode (the axion) and a scalar
mode of mass $\ms = \sqrt{2\la}\eta$.  During the phase transition, the
direction in field space is chosen at random in uncorrelated regions of the
universe, with the result that the field is forced to stay zero along lines
\cite{Kibble:1976sj}.  These lines form the cores of the axion strings
\cite{Davis:1986xc}.  The physical radius of the core is approximately 
\ben
\ws = \ms^{-1} = 1/\sqrt{2\la}\eta. 
\label{e:StrWid}
\een
In a flat Friedmann-Lema\^itre-Robertson-Walker (FLRW) metric, and well below the temperature of the phase transition, the dynamics of the field are well described by the classical equations of motion, which in conformal time and comoving coordinates are
\ben
\ddot{\Phi}+2\frac{\dot{a}}{a}{\dot\Phi}-\nabla^2 \Phi = -a^2 \lambda (\Phi^T \Phi-\eta^2)\Phi,
\label{eq:eom}
\een
where $a$ is the scale factor, dot denotes differentiation with respect
to conformal time $\tau$, and in the radiation era $a \propto \tau$.  
As part of the classical dynamics of the field, the axion strings evolve as relativistic strings, emitting both scalar and pseudoscalar modes.

The average energy-momentum of the system 
can be characterised by the total rest-frame string length and the root-mean-square (RMS) velocity of the network. 
These can be estimated from combinations of different energy density components \cite{Hindmarsh:2019csc,Hindmarsh:2021vih}. 
We showed in those papers that independent length and velocity estimators can be obtained from different combinations, all of them providing equivalent descriptions.  
In this paper we will use two such estimators for the comoving length of the strings: the ``winding'' length estimator $\Lwind$ and rest-frame string length estimator $\SlenRes$. These two measures are related to each other by the velocity, which can also be estimated from combinations of energy components.  
These are all defined in previous papers  \cite{Hindmarsh:2019csc,Hindmarsh:2021vih}, but we write them here again to keep the discussion self-contained.

The winding length estimator $\Lwind$ is  widely used. It is  a convenient measure on the lattice, defined as the number of plaquettes pierced by strings $n_\text{p}$ multiplied by the lattice spacing $\Delta x$. 
Whether a plaquette is pierced by a string is determined by the winding number of the vector $\Phi/|\Phi|$ around the plaquette. 
As strings do not pass orthogonally through the centre of the plaquette, various corrections can be applied; we follow the simplest, which is to compensate for the angle by multiplying by a factor 2/3 \cite{Fleury:2015aca}. 
This observable gives an estimate of the length of string in the FLRW or ``universe'' rest frame,
\ben
\Lwind = \frac{2}{3} n_\text{p} \Delta x\,.
\label{e:ellWinDef}
\een
We also study an estimate for the rest-frame string length, which is proportional to the total energy of the strings, including kinetic energy. In order to estimate the energy in strings, we weight components of the energy-momentum tensor so that strings contribute much more than the volume between them. 
The weighted kinetic, gradient and potential energy density components are 
\bea
 E_{\pi}  &=& \frac{1}{2} \int d^3 x \Pi^2 \WtFun(\Phi) \label{eq:EStrKin}, \\
E_{D} &=&  \frac{1}{2} \int d^3 x (\nabla\Phi)^2 \WtFun(\Phi)\label{eq:EStrGrad}, \\
E_{V} &=& a^2 \int d^3 x V(\Phi) \WtFun(\Phi)\label{eq:EStrV},
\eea
where $\Pi = \dot\Phi $.
Then, under the assumption that the field configuration is that of a straight static string boosted to a velocity $\dot\bX$, the field energy is proportional to the rest frame length $\SlenRes$,
\ben
E = E_{\pi} + E_{D}  + E_{V} = \mu_\WtFun ( 1 - f_{V,\mathcal{W}} \vAv^2 ) \SlenRes,
\label{e:TotEne}
\een
where $\vAv^2 = \vev{\dot\bX^2}$ is the mean square velocity, $\mu_\WtFun$ is the weighted energy per unit length, and $f_{V,\mathcal{W}}$ is the fraction contributed by the weighted potential energy density.   Note that the energy $E$ is a comoving quantity, coming from integrals over comoving coordinates of $-T_{00}$, the fully covariant 00 component of the energy-momentum tensor. The physical energy is equal to $a^{-1} E $.

The (comoving) Lagrangian is 
\ben
L = E_{\pi} -E_{D}  - E_{V} =  - \mu_\WtFun(1 - v^2) \SlenRes.
\een
Using these, the rest-frame string length in comoving coordinates can be estimated as
\ben
\SlenRes = \frac{E + f_{V,\mathcal{W}} L}{\mu_\WtFun(1 - f_{V,\mathcal{W}})} \, .
\label{eq:slenres}
\een
We choose the weight function to be the potential $\WtFun(\Phi)=V(\Phi)$, which is strongly peaked at the string cores. 
On the 3D lattice, the resulting constants $\mu_V$ and $f_{V,V}$  depend on the ratio of the lattice spacing to the comoving string core width, 
$\mu_V(\Delta x /\ws^c)$, $f_{V,V}(\Delta x /\ws^c)$, where $\ws^c = \ws/a(\tau)$. We outline our method for determining these functions in Appendix \ref{a:NumSolWtPar}.

The weighted energy and Lagrangian defined above also give an estimate of the mean squared (peculiar) velocity of string
\ben
v^2_L = \frac{E - L}{E + f_{V,V} L} .
\label{eq:vLagWt}
\een
The mean squared velocity can also be extracted from the ratio of kinetic to gradient energy in the scalar field $R_{\text{s}} = {E_{\pi}}/{E_{D}}$, 
which gives the scalar velocity estimator, 
\ben
\vAv_{\text{s}}^2 = \frac{2R_{\text{s}} }{1+ R_{\text{s}} }.
\label{eq:vel_s}
\een
In previous work, we use the scalar velocity estimator. While both measures agree very well,  $\vAv_{\text{s}}$ was observed to fluctuate a little less than $v_L$. In our current simulation, there is very little difference, and we use $v_L$ for consistency.

Finally, 
the length estimator derived from counting plaquettes is related to the rest frame estimator and the local Lorentz factor $\gamma$ by  
 \ben
 \Lwind=\SlenRes\left\langle\gamma^{-1}\right\rangle\,.
 \label{rela}
 \een
Most other groups study only the universe frame (winding) length estimator, which 
evolves differently from the rest frame length due to the velocity evolution.

\section{String scaling and modelling}
\label{sec:scaling}

Theoretical arguments and numerical simulations (see \cite{Vilenkin:2000jqa,Hindmarsh:1994re,Hindmarsh:2011qj} for reviews) show that strings, originating from both global and  gauge symmetry breaking, are formed in a tangle consisting (in an infinite universe) of an infinitely long string and a distribution of loops, which are all random walks on large scales. The step length of the random walk is set by the correlation length of the field at the phase transition, determined by the Kibble-Zurek mechanism for a continuous phase transition \cite{Kibble:1976sj,Zurek:1996sj}, the mean bubble spacing for a first order transition \cite{Hindmarsh:1993av}, or the thermal correlation length at a cross-over in a gauge theory \cite{Hindmarsh:2001vp,Hindmarsh:2000kd,Blanco-Pillado:2007ihs}.

Once formed, the axion strings start moving under acceleration by curvature, 
and decay into scalar and axion radiation, resulting in the rapid disappearance of loops and the dilution of strings. 
The length per unit volume of string decreases, and the mean string separation $\xi$ increases. The mean string separation in physical units is defined from the comoving string length $\ell$ per unit volume $\vol$ as
\ben
\frac{1}{\xi^2} = \frac{\ell}{a^2\vol}.
\label{e:xiDef}
\een

It is convenient to express the mean string separation in terms of a variable 
\bea
x &=& \frac{\xi}{t}, \label{e:xDef}\,
\eea
where $t$ can either be taken to be the cosmic time or the time since the phase transition.  The former is more convenient in cosmological simulations, while the latter is more convenient in Minkowski space \cite{Smith:1987pv,Vincent:1996rb,Vincent:1997cx}.  
Equivalently, one can consider the scaled (physical) length density
\bea
\zeta &=& \frac{t^2\ell }{a^2\vol}=\frac{1}{x^2} . \label{e:zetaDef}
\eea
Substitution of $\Lwind$ or $\SlenRes$ then leads to rest-frame and universe frame (Lorentz contracted) versions of the length density parameter, $\zer$ and $\zew$, respectively. We remark that $\SlenRes$ and hence $\zeta_r$ are not computed from $\Lwind$, so a useful consistency check arises from the relation $\zer \simeq \zew \gamma(v)$, where $\gamma(v)$ is the Lorentz factor computed from the RMS velocity $v$. 
Some authors \cite{Klaer:2017ond,Klaer:2017qhr,Klaer:2019fxc} define a length density parameter in terms of comoving length and conformal time, $\zeta^\text{c} = \tau^2 \ell/\vol$.  In the radiation era, $\zeta^\text{c} = 4\zeta$.

The evolution of $x$ and the RMS velocity $v$ is controlled by the complicated interactions of the strings with each other and with the fields they are made from.  
The fundamental hypothesis behind scaling is that as the average radius of curvature of the strings gets large compared with the string width, the string width may be neglected in the evolution,
and the only velocity scale is that of waves of the string, in this case the speed of light. 
The string equations of motion can be used to derive models of the joint evolution of the mean curvature radius and the RMS velocity, called velocity one-scale (VOS) models \cite{Kibble:1984hp,Martins:1996jp,Martins:2000cs}. 
Uncertainties about the averaging of products of dynamical variables and about the energy loss mechanism mean there are are several such models.

In a scaling system, the mean string separation is of order the curvature radius, 
there is no explicit dependence on the string width, 
and the general form for VOS models is an autonomous dynamical system  
\bea
\frac{d x}{d \La} &=& f_x(x, v) ,
\label{e:AutDynSysx} \\
\frac{d v} {d\La} &=& f_v(x, v),
\label{e:AutDynSysv}
\eea
where $\La = \ln(t/t_0)$, and $t_0$ is an arbitrary time scale. 
Such systems are characterised by null clines $f_x(x, v) = 0$, $f_v(x, v) =0$ and fixed points $(x_*, v_*)$ where null clines intersect.  

In Ref.~\cite{Hindmarsh:2021vih} we showed that the evolution of $\xr$ (the $x$ variable (\ref{e:xDef}) derived from the rest-frame length density) and $\vs$, when analysed in the context of a simple two-parameter VOS model along the lines outlined above, was consistent with a spiral approach to a fixed point, 
with $v_{{\rm s},*} = \vsStarVOS(\dvsStarVOS)$ and $x_{{\rm r},*} = \xStarRestVOS(\dxStarRestVOS)$, corresponding to 
$\ze_{{\rm r},*} =\zzRestMeansVOS(\zzRestErrsVOS)$ and 
$\ze_{{\rm w},*} =\zzWindMeansVOS(\zzWindErrsVOS)$.

That the approach should be a spiral is a feature of the VOS model used in \cite{Hindmarsh:2021vih}, which is a simplification of the original \cite{Martins:1996jp,Martins:2000cs}. In general, the trajectories of VOS models need not execute a spiral approach, as seen for example in Fig.~1 of \cite{Correia:2021tok}. 
An accurate VOS model for axion string evolution over the whole region of phase space explored by simulations remains to be constructed. 

A notable feature of the dynamics is an approximately logarithmic growth $\zeta \propto \La$ at $\zeta \lesssim 0.5$, with slope around $0.2$ \cite{Gorghetto:2018myk,Kawasaki:2018bzv,Gorghetto:2020qws,Buschmann:2021sdq,Saikawa:2024bta} termed the ``attractor solution'', which is proposed as the true long-term behaviour \cite{Gorghetto:2018myk,Gorghetto:2020qws}.
One way to model this behaviour is through invoking an explicit (but weak) dependence on the string width in one or more of the parameters of the VOS equations. Such a scale dependence is argued to exist in Ref.~\cite{Gorghetto:2020qws} and incorporated into models describing the string density evolution in Refs.~\cite{Klaer:2019fxc,Buschmann:2021sdq}. 

In the context of standard scaling, the growth is viewed as a low-density transient.  That such transients exist and can be fitted with a power of $\La$ \cite{Gorghetto:2018myk,Gorghetto:2020qws} is to be expected given the structure of the equations (\ref{e:AutDynSysx}, \ref{e:AutDynSysv}) \cite{Hindmarsh:2021vih}.  The form of the transient is a clue to the structure of the equations at low $\zeta$, and hence large $x$.

Note that the RMS velocity decreases during the low-density growth phase
(see Fig.~7 of \cite{Gorghetto:2020qws} and Fig.~9 of \cite{Kim:2024wku}), which is consistent with it being a transient.  The question of the long-term value of the RMS velocity has not been addressed in the long-term growth scenario. It should also be noted that the original claims of long-term logarithmic growth in $\ze$ were based on simulations which ended with $\ze_{{\rm w}} < \ze_{{\rm w},*}$, and the largest recent simulations still have not significantly exceeded $\ze_{{\rm w},*}$.

\section{Simulations}
\label{sec:simulations}

We simulate the dynamics of the system by evolving a discretized version of (\ref{eq:eom}) in cubic periodic lattices with spacing $\Dx$.  The Laplacian has a 7-point stencil of nearest neighbours, and the time evolution is a generalised leapfrog
\bea
\Pi^{i+\frac{1}{2}} &=& \Pi^{i - \frac{1}{2}} \left(\frac{a^{i - \frac{1}{2}}}{a^{i + \frac{1}{2}}} \right)^2  
\!\!\! + F[\Phi^i, a^i] \left(\frac{a^i }{a^{i + \frac{1}{2}}} \right)^2 \!\! \Delta \tau, \label{e:SchPi}\\
\Phi^{i+1} &=& \Phi^{i} + \Pi^{i+\frac{1}{2}} \Delta \tau,
\label{e:SchPhi}
\eea
where $i$ labels the time, $\Delta \tau$ is the timestep, and $F$ is the force term containing the Laplacian and the derivative of the potential. Explicitly, 
\bea
 F[\Phi^i, a^i]  &=& \sum_{\hat{\bf e}}\left[ \Phi^i(\bx + \hat{{\bf e}} ) - 2 \Phi^i(\bx) +   \Phi^i(\bx - \hat{{\bf e}} ) \right]/\Dx^2 \nonumber \\
&&  + (a^i)^2 \lambda \left[(\Phi^T\Phi)^i(\bx) - \eta^2 \right]\Phi^i(\bx),
 \eea
where $\hat{\bf e} \in \{\hat\bx, \hat{\by}, \hat{\bz}\}$ are the lattice unit vectors and $\bx$ the lattice site.  
We drop the lattice site notation in Eqs. (\ref{e:SchPi}, \ref{e:SchPhi}) for compactness.

This scheme has a time reversal symmetry $i+ \frac{1}{2} \leftrightarrow i -  \frac{1}{2}$, $i+1 \to i - 1$. This is a generalisation of a feature of the leapfrog scheme.

We fix code units by choosing $\eta = 1$. In these units, the minimum energy state has $|\Phi| = 1$, the lattice spacing and timestep are $\Dx=0.5$ and $\Delta \tau=0.1$ respectively.  The choices of lattice spacing and timestep satisfy the Courant-Friedrichs-Levy (CFL) condition for the hyperbolic system. Total energy is covariantly conserved to within 5\% in all simulations.  

Since we use periodic boundary conditions, there is an upper limit in the dynamical range of the simulation of half a light-crossing time, beyond which it is possible that  finite volume effects appear. We were able to increase the dynamical range comparing to the range in previous works \cite{Hindmarsh:2019csc,Hindmarsh:2021vih}, by performing simulations with up to 16k lattice sites per side, where $n$k means a side length of $1024n$ sites.

The field configuration is initiated as follows: At conformal time \tStart\, we set  
$\Pi$ to zero everywhere, and $\Phi$ to be a Gaussian random field
with power spectrum   
$$P_\Phi(k)= Ae^{-\half\left(k\IniCorLen\right)^2}\,,$$ 
where $A$ is chosen so that 
$\langle\Phi^2\rangle=1$ and  $\IniCorLen$  is the field correlation
length in comoving coordinates. For this work we have used values of $\IniCorLen \in \{5,10,20,40,80\}\,,$ and for 16k simulations also $\IniCorLen  \in \{160,320\}$.  Different values of $\IniCorLen$ create initial configurations with different string densities: the higher  the $\IniCorLen$ the lower the initial string density. 
In order to remove any excess energy in the initial configuration we allow this configuration to evolve with a diffusion equation with unit diffusion constant until conformal time \tDiff. 
During the diffusion period,  the scale factor is kept constant and the time update is
\bea
\Pi^{i+\frac{1}{2}} &=& F_n[\Phi^i]  , \\
\Phi^{i+1} &=& \Phi^{i} + \Pi^{i+\frac{1}{2}} \Delta \tau',
\eea
where $\Delta \tau'$ is chosen to satisfy the CFL condition for the diffusion equation.
 
Another issue that needs attention is that by the end of the simulation we need the strings to be well resolved (in comoving coordinates), which means that at early times the string widths are very big, easily bigger than the Hubble length, which also means that the relaxation of the field to the string solution would be very slow. To avoid such issues, we promote the self-coupling $\lambda$ to be a time-dependent parameter \cite{Press:1989yh,Bevis:2006mj}
$$\lambda=\frac{\lambda_0}{a^{2(1-s)}}\,,$$ 
where $\lambda_0$ is a constant throughout the simulation, taken to be 2, and $s$ takes different values in different periods of the simulation.
In the period where we take data, $s=1$, the physical value. We normalise the scale factor so that $a=1$ at the end of the simulation, at conformal time $\tEnd$, which is taken to be a little larger than half the light-crossing time of the simulation volume, or $n \Dx/2$.

We also arrange that the comoving string width during the diffusive evolution is 
equal to their final comoving width. This minimises the time taken to relax to smooth field configurations containing strings. 
At the end of the diffusion period $\tDiff$ the comoving string width is therefore much smaller than should be, given the ratio of conformal times $\tDiff/\tEnd$. In order to bring the string width to the correct value, 
we set $s=-1$ until a conformal time $\tcg$,  which is when the string core
has expanded to its physical width. It is then that the physical evolution of the system starts, and we regard the state of the system at this time to be the initial condition. We note that the evolution between $\tDiff$ and $\tcg$ is totally self-similar, in that the string width remains a constant fraction of the horizon \cite{Klaer:2019fxc}.

By appropriate choice of $\IniCorLen$ we can simulate networks with a spread of string densities around  the expected final state $\zer \simeq 1$ at the start of the physical evolution. Here we highlight the importance of having as wide a range as possible to investigate the effect of the initial string density on the final time $\tEnd$.

Table~\ref{table} shows the parameters $\IniCorLen$, $\tcg$ and  $\tEnd$  
used for the initial conditions and subsequent evolution of the network, and the number of runs at each parameter set, each with a different seed for the random number generator. In Table \ref{t:ZetIni} we show the values of the string density parameter $\zer$ at the end of core growth evolution for each choice of initial correlation $l_\phi$ length and box size.  In other words, this is the value of $\zer$ at the beginning of the physical evolution of the system. All cases have starting time $\tStart=50$ and $\tDiff=70$. 

 \begin{table}[h!]
\begin{tabular}{ c | c | c | l | c | c |}
 Box size & Seeds & {\hfill  Runs \hfill} & {\hfill $\IniCorLen$ \hfill} & \tcg & \tEnd   \\ \hline
4k &  9    & 45 & 5,10,20,40,80 & 271 & 1050  \\
8k & 9     & 45 & 5,10,20,40,80 & 383  & 2100 \\
12k & 6 & 30 & 5,10,20,40,80&470 &3150 \\
16k & 8   & 56 & 5,10,20,40,80,160,320 &539  & 4150  \\
\end{tabular}
 \caption{\label{table}
 Parameters $\IniCorLen$ (initial field correlation length), $\tcg$ (conformal time at the start of physical evolution) and  $\tEnd$ (conformal time at the end of the evolution) and number of simulations used to generate string networks. A box size $n$k means that the number of lattice sites per side of the cubic simulation volume is $1024n$. All values of conformal \tcg\ correspond to physical time $t_\text{cg}=70/\ms$, while conformal time  \tEnd corresponds to physical time $t_\text{end} = \tEnd/2$.}
\end{table}

\begin{table}[h!]
\begin{tabular}{ l | l | l | l | l }
{\hfill $\IniCorLen$ \hfill}  & {\hfill 4k \hfill} &  {\hfill 8k \hfill} &  {\hfill 12k \hfill} &  {\hfill 16k \hfill} \\
\hline
5 & $ 2.275(4) $  & $ 2.063(6) $  & $ 1.977(2) $  & $ 1.938(3) $  \\
10 & $ 1.856(3) $  & $ 1.757(5) $  & $ 1.723(3) $  & $ 1.694(2) $  \\
20 & $ 1.407(4) $  & $ 1.411(4) $  & $ 1.418(3) $  & $ 1.415(2) $  \\
40 & $ 0.922(4) $  & $ 1.035(4) $  & $ 1.101(5) $  & $ 1.133(1) $  \\
80 & $ 0.472(5) $  & $ 0.607(3) $  & $ 0.706(1) $  & $ 0.7753(9) $  \\
160 &              &              &              & $ 0.404(1) $  \\
320 &              &              &              & $ 0.162(2) $  \\
\end{tabular}
 \caption{\label{t:ZetIni}
Values of the string density parameter $\zer$ at the start of physical evolution $t_\text{cg} = 70/\ms$, for different initial field correlation lengths $\IniCorLen$ and box sizes. 
 }
\end{table}

 \begin{figure*}[p]
    \centering
    \includegraphics[width=0.85\columnwidth]{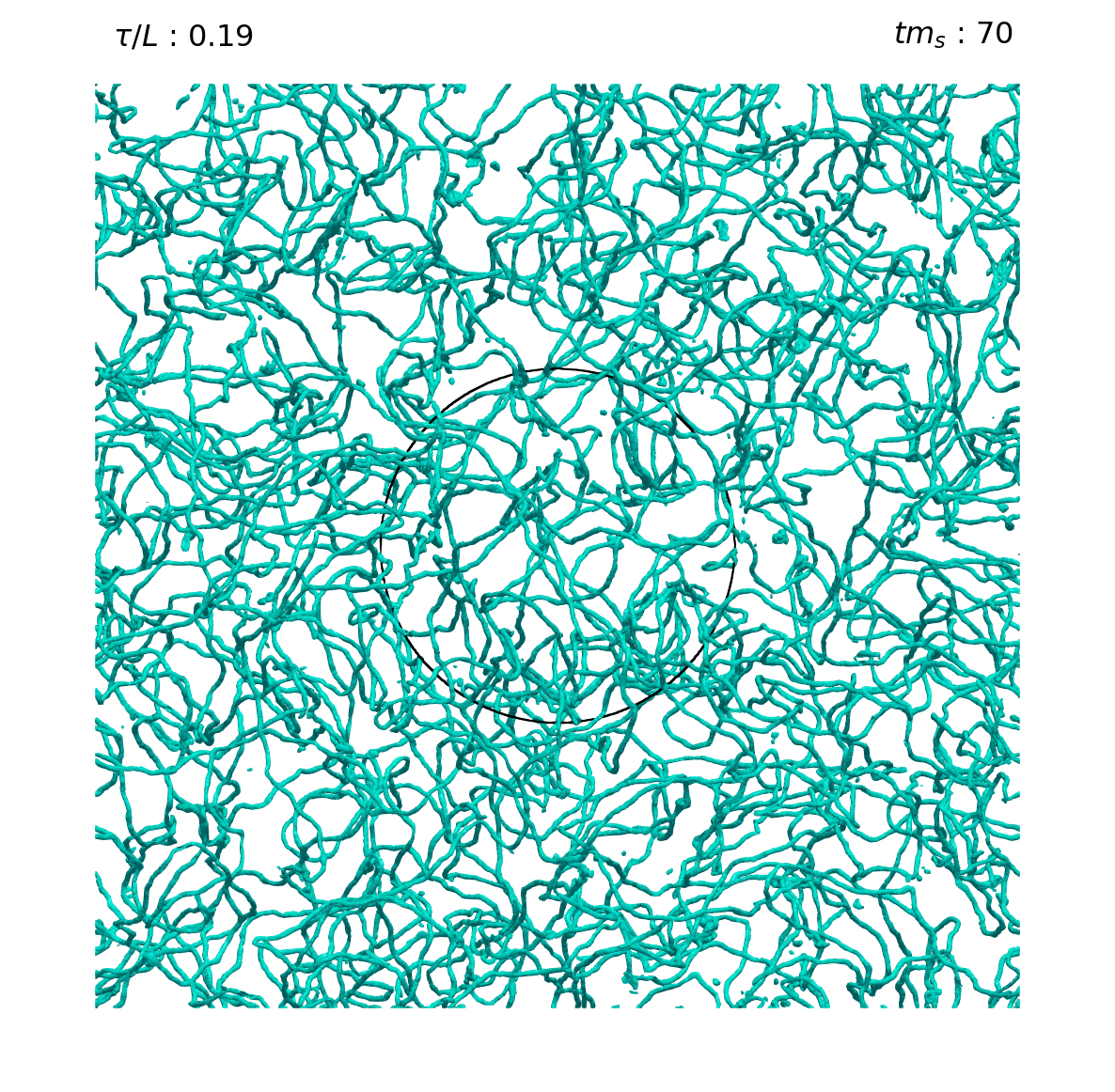}
    \includegraphics[width=0.85\columnwidth]{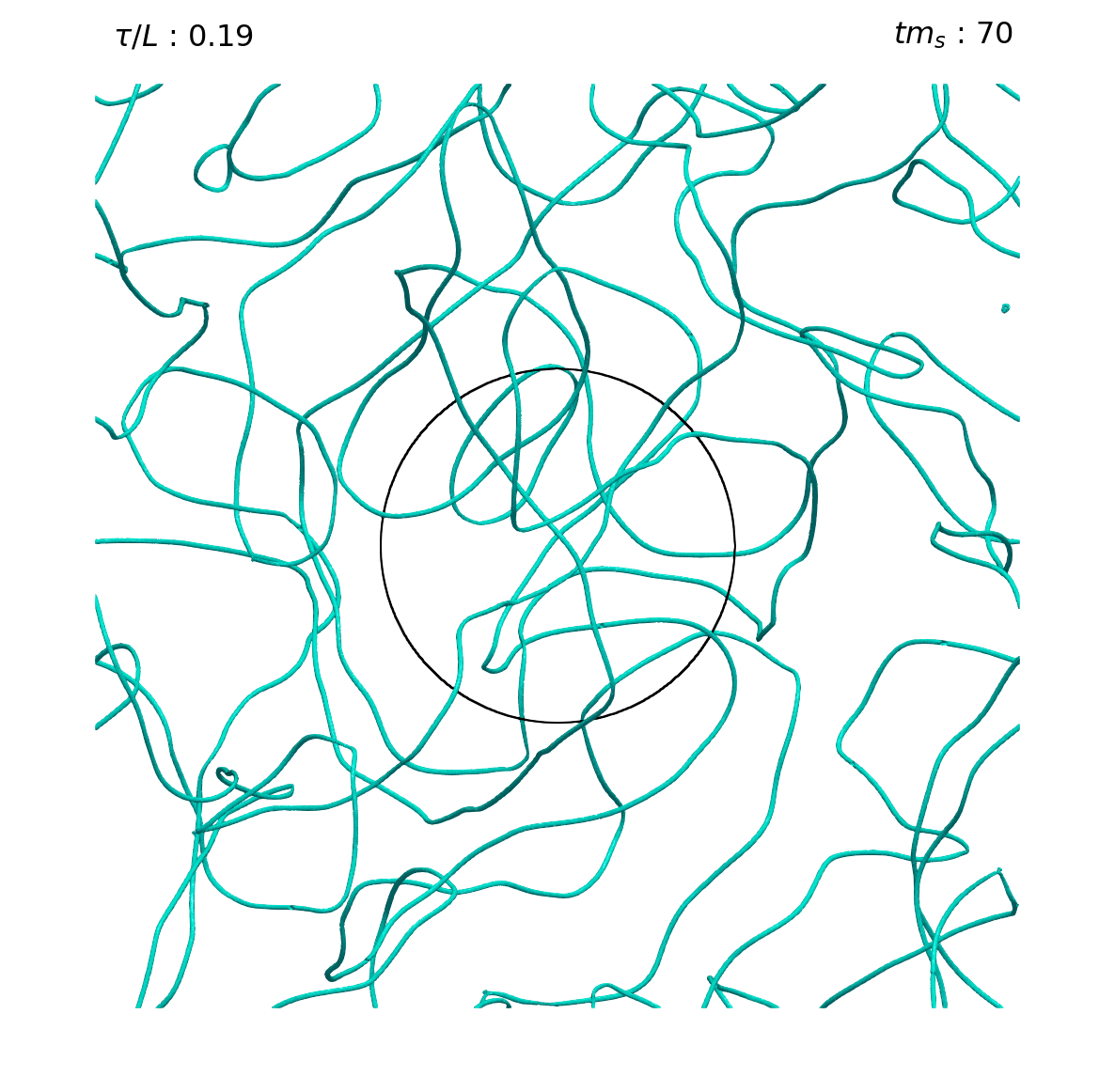}
    \includegraphics[width=0.85\columnwidth]{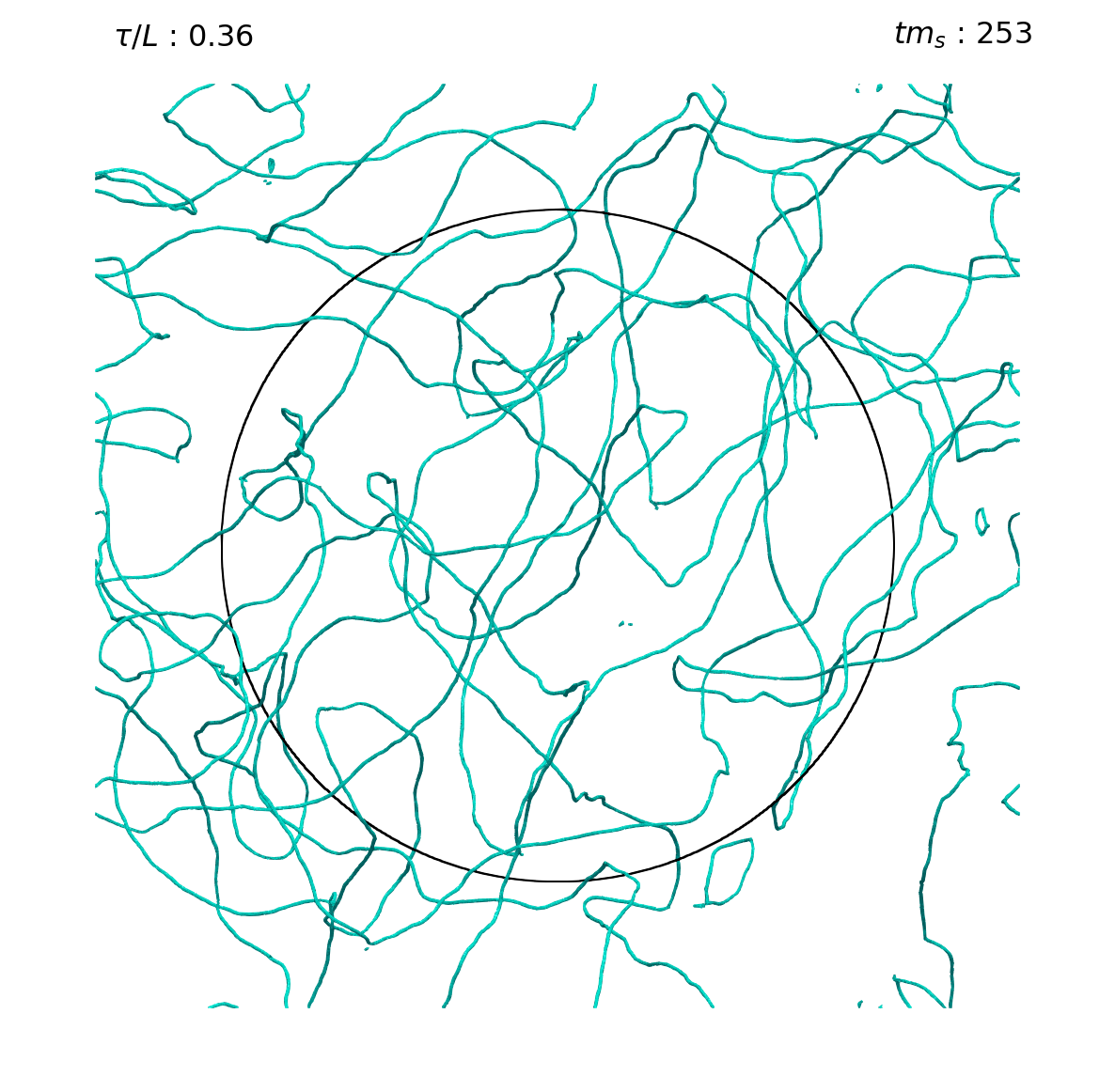}
    \includegraphics[width=0.85\columnwidth]{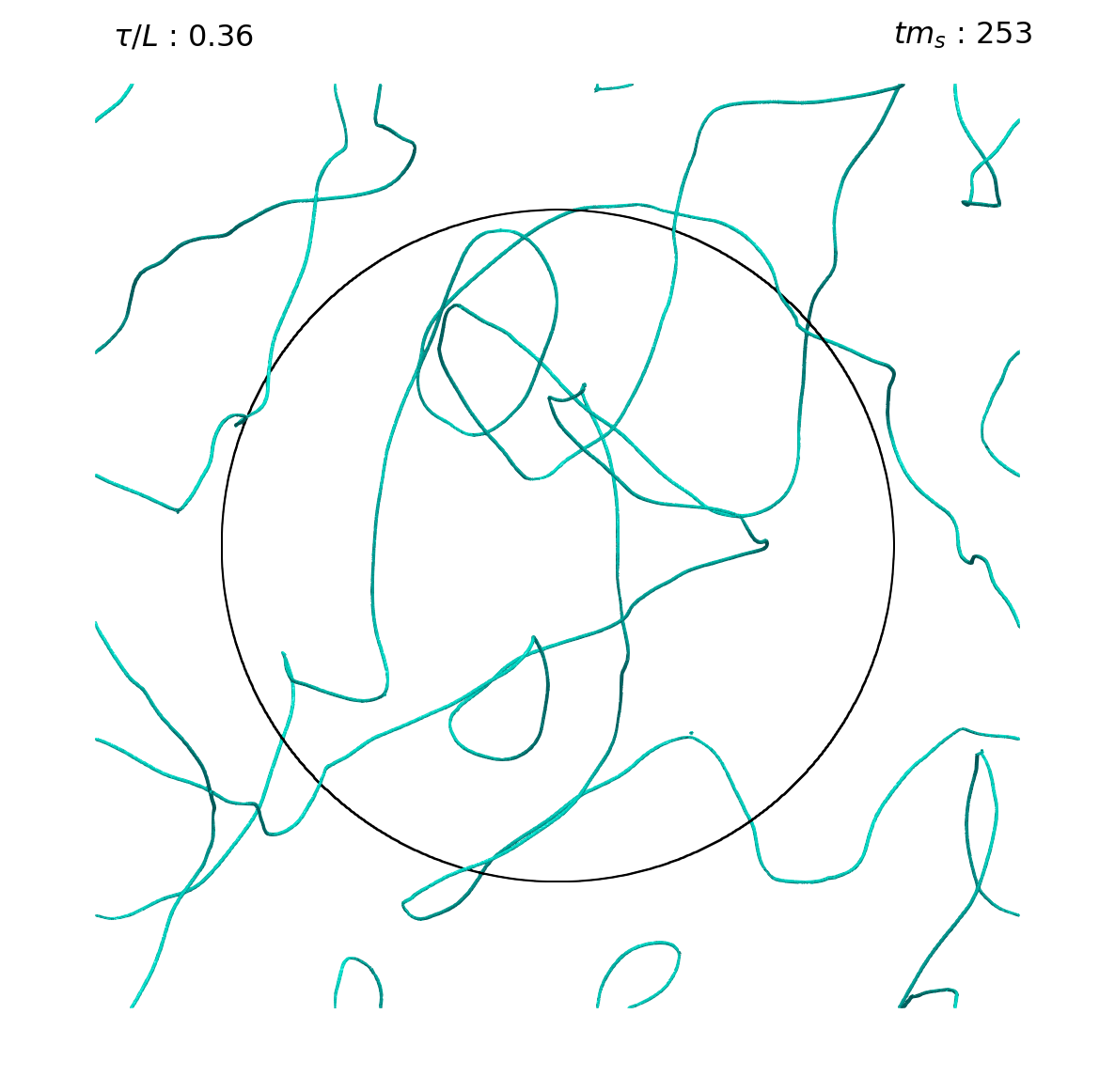}
    \includegraphics[width=0.85\columnwidth]{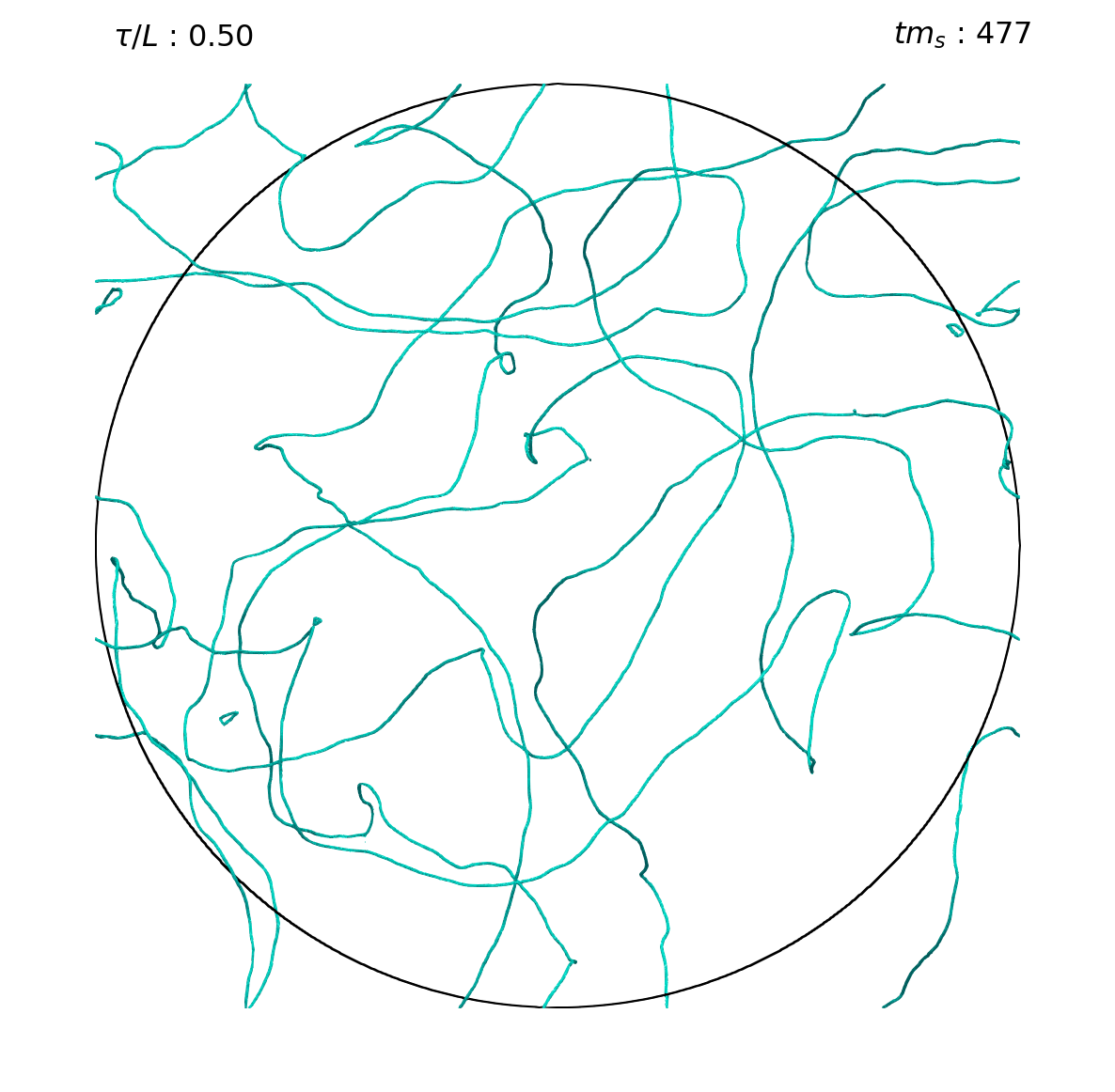}
    \includegraphics[width=0.85\columnwidth]{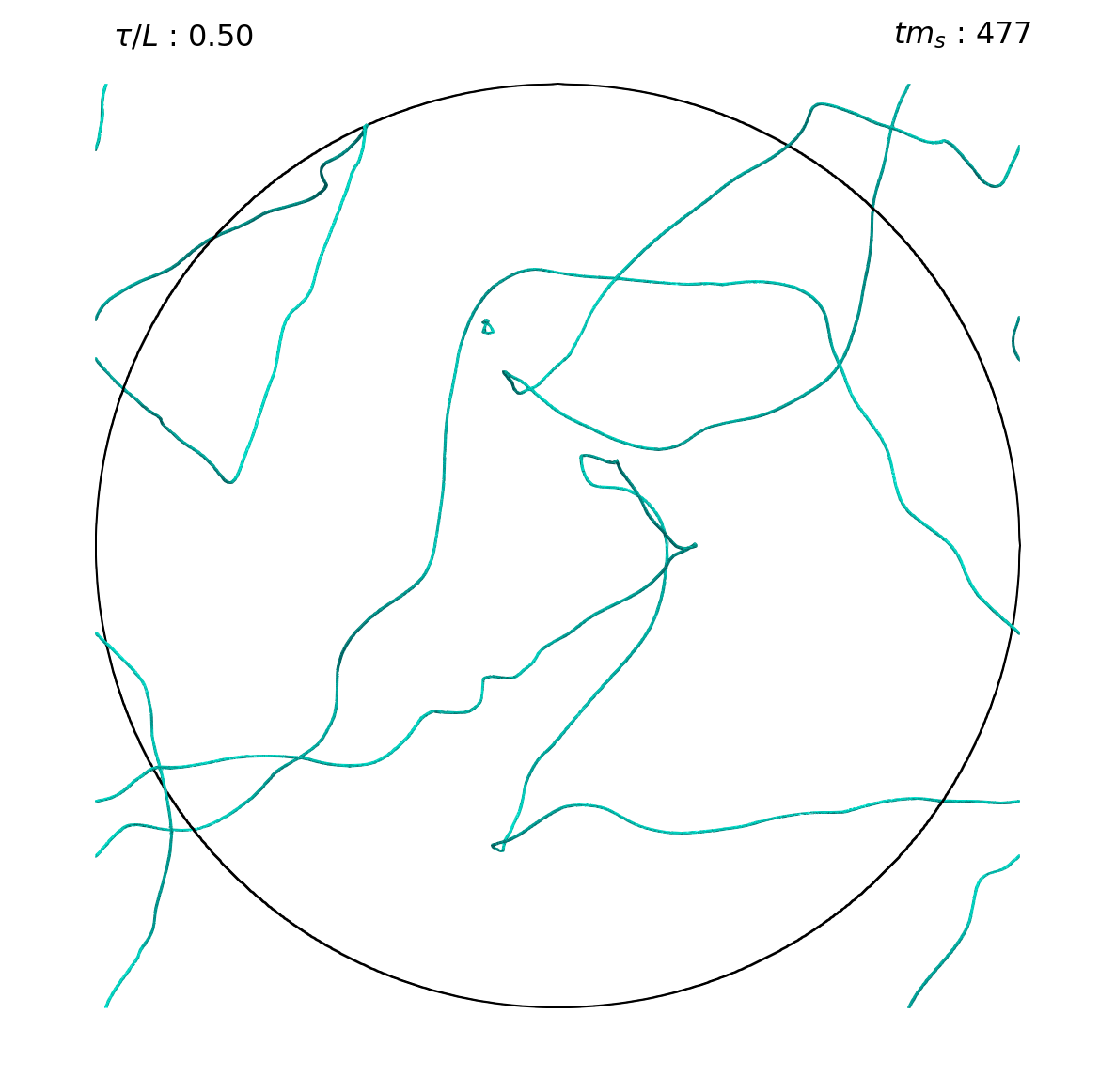}
    \caption{\label{fig:visualization} Visualisation of $|\Phi| = 0.5$ isosurfaces at three different conformal times (given in units of box side length, $\tau/L$, where $L=1024$) or equivalently in cosmic time (multiplied by scalar mass $tm_s$) for two initial field correlation lengths $l_\phi$ in radiation epoch. We display $l_\phi=5$ on the left-hand-side panels, and on the right-hand-side ones  $l_\phi=80$. The top panel corresponds to an early time close to the beginning of physical evolution ($\tau/L=0.19$ or $t\ms=70$), the middle panel to an intermediate conformal time ($\tau/L=0.36$ or $t\ms=253$), and the last panel to half-a-light time crossing ($\tau/L=0.5$ or $t\ms=477$). The black circle has a radius set to conformal time $\tau$, equal to the comoving Hubble length in this radiation era simulation. The full animations can be found at the following location: \cite{correia_2024_14160869}.} 
 \end{figure*}

In Fig.~\ref{fig:visualization} we show isosurfaces of the scalar field magnitude $|\Phi| = 0.5$ from a 2k simulation with initial field correlation lengths $\IniCorLen = 5$ and $\IniCorLen = 80$, at conformal times $\tau/L = 0.19\, , 0.36\, , 0.50$, where $L = 1024$ is the box side length in comoving distance units. Also given are the corresponding physical (cosmic) times $ t = a(\tau) \tau/2$.
The circle is a section of the light cone of a point at the centre of the box at $\tau=0$. It has radius $\tau$, equal to the comoving Hubble radius in this radiation era simulation. 

Evident in the simulation is a convergence from a greatly differing initial number of causal lengths $\tau$ of string per causal comoving volume $\tau^3$ towards similar values. 
The 2k simulations are too small to show statistically indistinguishable states at $\tau/L \simeq 0.5$, while the 16k simulations (which do show convergence for these initial conditions, as we demonstrate in the next section) are impractical to visualise in this way.

\section{Results: standard scaling analysis}
\label{s:ResStaScaAna}

This section contains the main results of the campaign of simulations presented in this work. We focus on two globally averaged quantities: the mean string length $\ell$, in physical coordinates, and the RMS velocity $v$.  
We estimate them as discussed in Section \ref{sec:ModelSims}. 
We recall that there are two different string lengths to consider: the length of string in universe frame $\ell_\text{w}$, which is Lorentz contracted, and the length of string in its own local rest frame $\ell_\text{r}$, which is proportional to the total energy in the string. From these we construct three different measures of the density of string in the simulation:
the mean string separation $\xi$ \eqref{e:xiDef}, 
a dimensionless mean string separation $x$ \eqref{e:xDef} and
the dimensionless length density parameter $\ze$ \eqref{e:zetaDef}.

\subsection{Time series data}
 
\begin{figure}[h]
    \centering
    \includegraphics[width=\columnwidth]{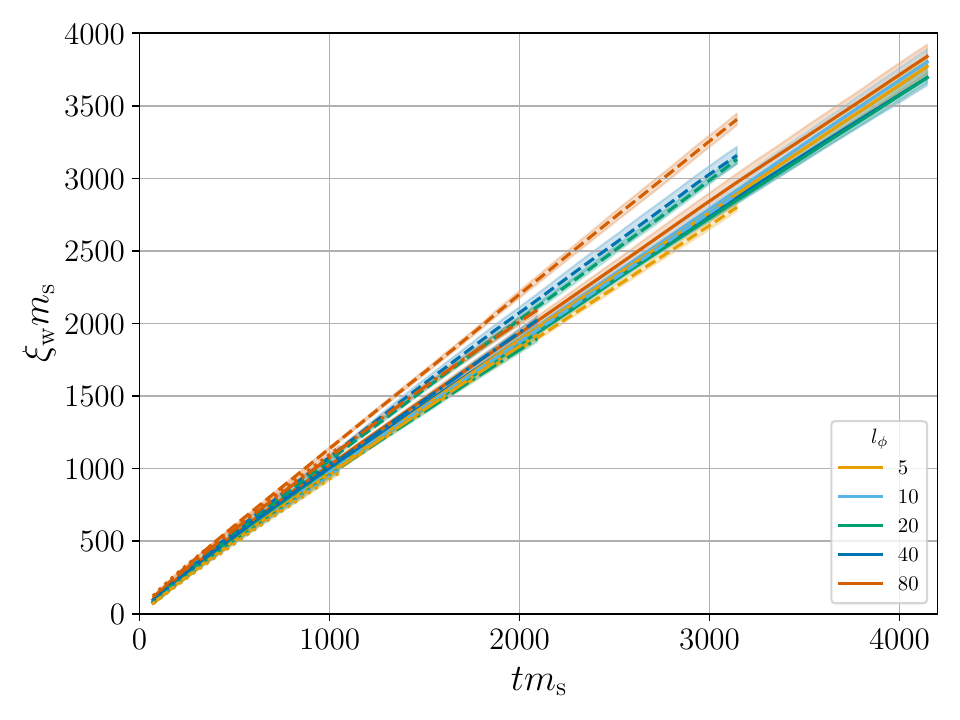}
    \includegraphics[width=\columnwidth]{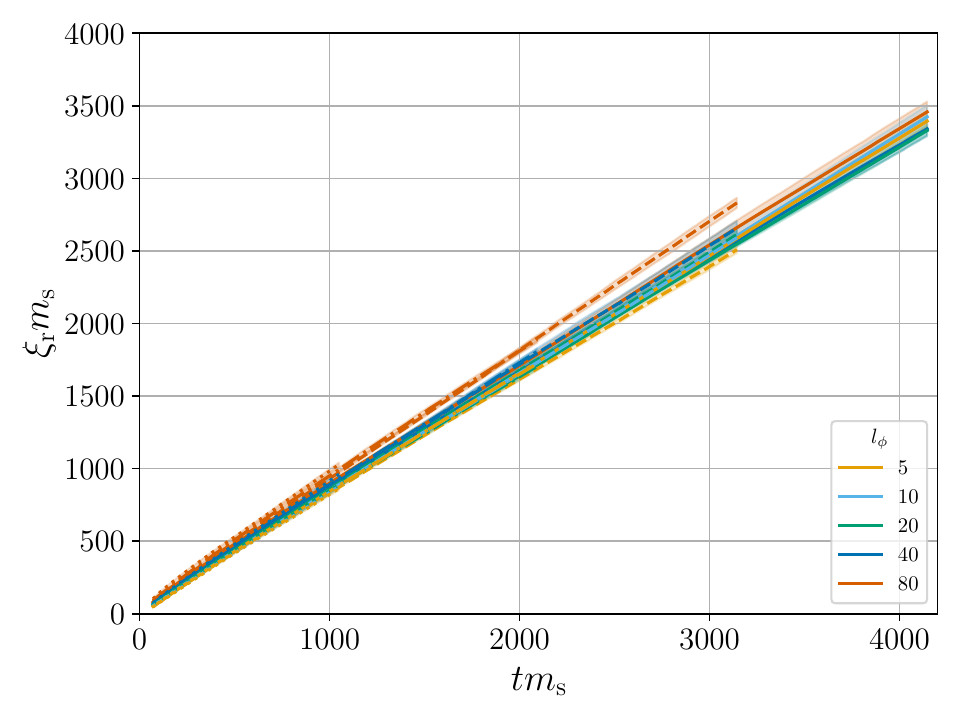}
    \caption{\label{fig:xiScaling}  Mean string separations $\xi$  for simulations with correlation lengths $\IniCorLen\eta=5,10,20,40,80$, for 4 different box sizes: 4k (dotted), 8k (dash-dotted), 12k (dashed) and 16k (continuous). 
 Top:    $\xiw$, derived from the universe-frame string length computed using the winding length estimator \eqref{e:ellWinDef}.  Bottom:  $\xir$ as computed using the rest-frame length estimator \eqref{eq:slenres}. The shaded regions show the $1\sigma$ errors on the mean. 
}
 \end{figure}

Figure~\ref{fig:xiScaling} shows the mean string separations computed from the string length in the universe and string rest frames $\xiw$ and $\xir$, plotted against cosmic time. The figure shows the simulations for five initial field correlation lengths $\IniCorLen =5,10,20,40,80$,  and also four box sizes, i.e., 4k (dotted), 8k (dash-dotted), 12k (dashed) and 16k (continuous). The different box sizes can also be identified by the time where the line stops, with the shortest lines corresponding to 4k  simulations and the longer ones to 16k. The solid lines show the mean values over runs and the shaded regions are the $1\sigma$ errors on the mean.
The data starts at the start of the physical evolution, at conformal time $\tcg$, corresponding in all cases to physical time $t_\text{cg} = 70/\ms$.

The mean string separation show a late-time behaviour qualitatively compatible with a linear growth, as expected in the standard scaling.

\begin{figure}[h]
    \centering
    \includegraphics[width=\columnwidth]{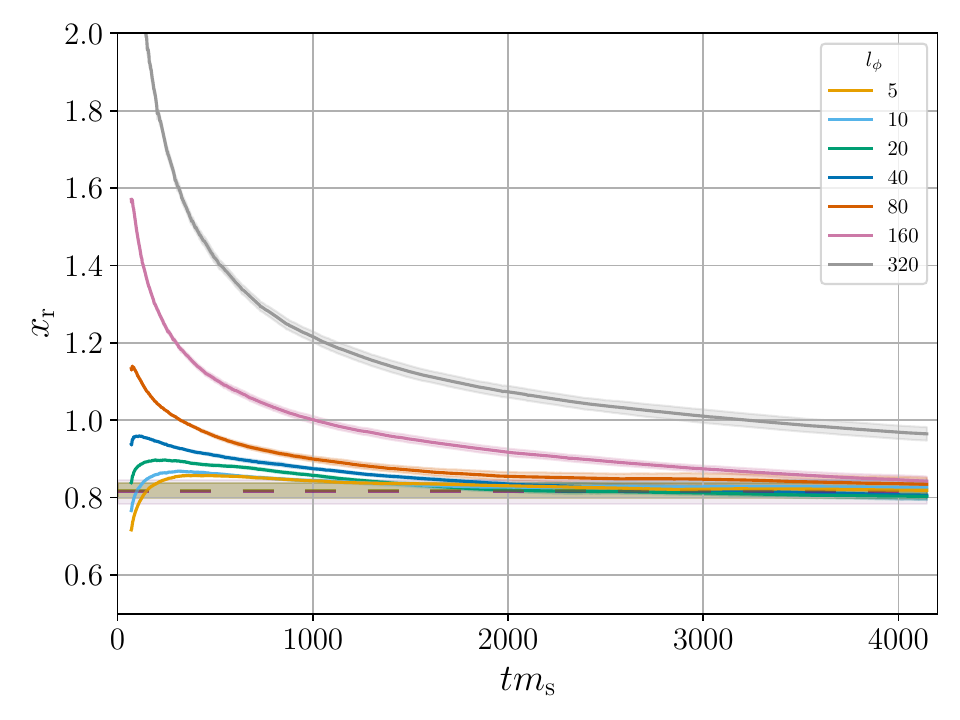}
    \caption{\label{fig:xScaling}  Plot of $x_{\rm r}$ (\ref{x})  for  16k simulations, for all initial correlation lengths ($\IniCorLen\eta=5,10,20,40,80,160,320)$. The horizontal dashed lines and shaded bands correspond to the asymptotic mean values and $1\si$ errors obtained in \cite{Hindmarsh:2021vih} (purple) and this work (olive green). Note that both estimates lie on top of each other.
}
 \end{figure}

We now look at the scaling of the network  through the 
scaled mean string separation \eqref{e:xDef}.  Here we focus on the scaled string separation derived from the rest-frame length 
\ben
  x_r=\xir/t\,.\label{x}
\een
Fig.~\ref{fig:xScaling} shows this variable in the biggest simulation boxes (16k), where the set of initial field correlation lengths extends to larger values 
($\IniCorLen=5,10,20,40,80,160,320$) and therefore there are even more widely separated (less dense) strings in the initial state. Horizontal dashed lines and shaded bands correspond to the asymptotic mean values and $1\si$ errors obtained in \cite{Hindmarsh:2021vih} (purple) and this work (olive green). Note that both estimates lie on top of each other.
The plot shows how a wide variety of different initial string separations (relative to $t$) tend to the similar values for $x_r$, at around 0.8. 
The values of $\xr$ in simulations with the highest $\IniCorLen$ (320) are about 20\% higher at the end of the simulations than the rest, but are still decreasing.  

\begin{figure}[h]
    \centering
    \includegraphics[width=\columnwidth]{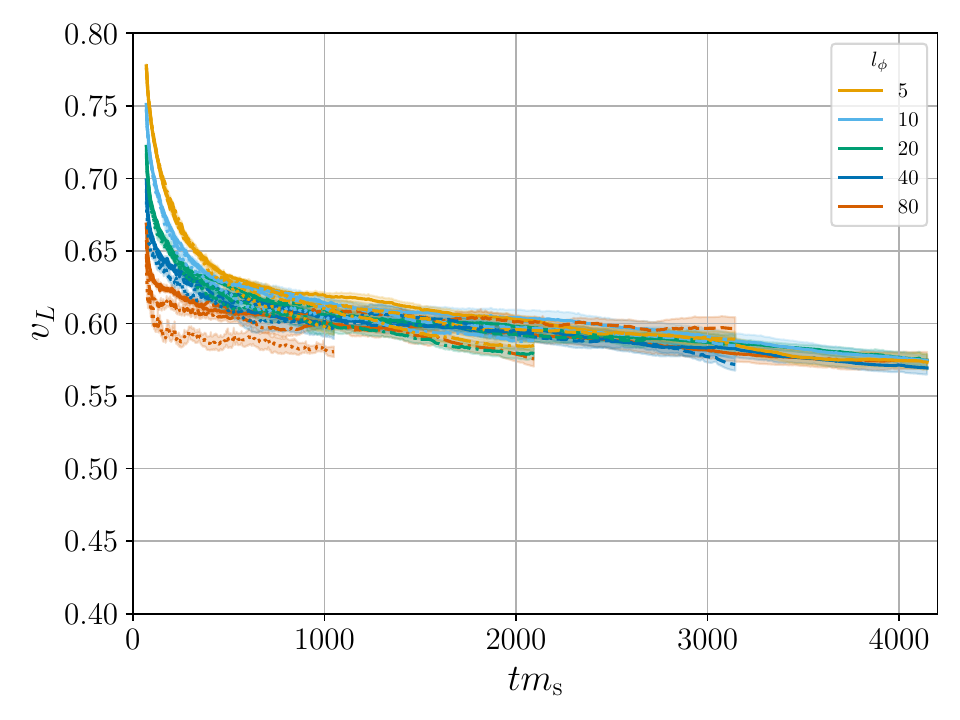}
    \caption{\label{fig:vScaling}  Plot of the root mean square velocity  (\ref{eq:vel_s}) for all simulations. The color coding  follows that of Fig.~\ref{fig:xiScaling}.
}
 \end{figure}

We now turn to  the  RMS velocity of Eq.~(\ref{eq:vel_s}).   Fig.~\ref{fig:vScaling} shows   $v$   for the different simulations at all volumes.

The RMS velocity estimator also shows an evolution consistent with the standard scaling picture, which predicts a slow approach to a constant value, as outlined in Section \ref{sec:scaling}. 
The constant value appears to be a little below $v \simeq 0.6$.  At early times, and for low initial field correlation lengths, the approach is from above. As mentioned earlier, the higher value of the RMS velocity at early times is a result of the initial acceleration from rest, and causes the universe-frame string length estimator (derived from plaquette-counting) to be significantly smaller than the rest-frame length estimator.

\begin{figure}[h]
    \centering
    \includegraphics[width=\columnwidth]{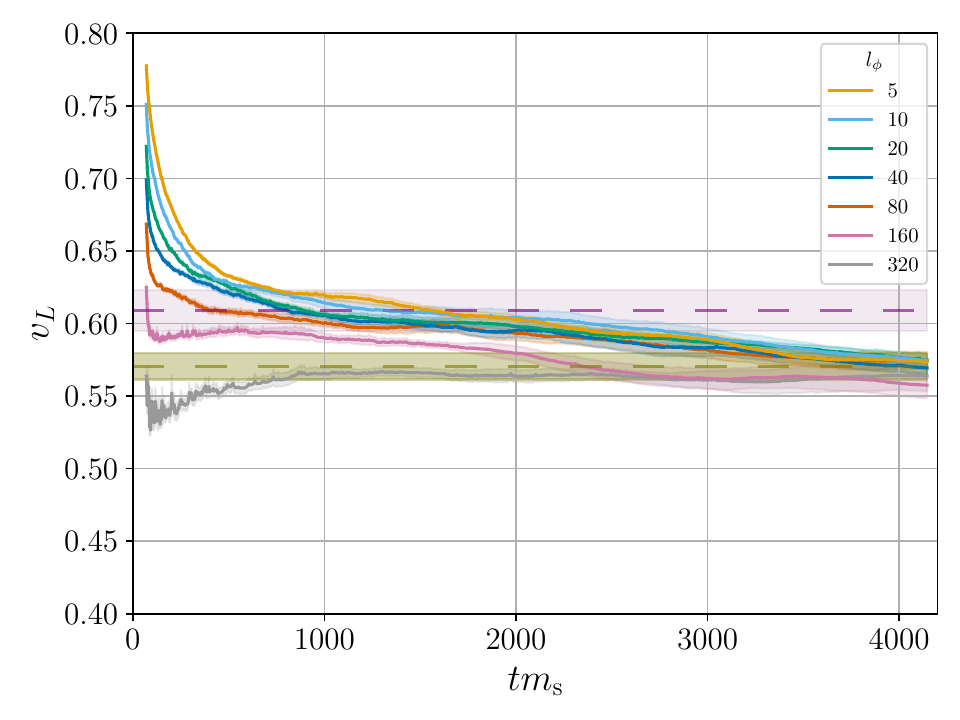}
    \caption{\label{fig:vScaling16k}  Plot of RMS velocity $v_{L}$ (\ref{eq:vLagWt}) for  16k simulations, with all initial correlation lengths $\IniCorLen$. The horizontal dashed lines and shaded bands as in Fig.~\ref{fig:xScaling}.
}
 \end{figure}

In Fig.~\ref{fig:vScaling16k} we show the RMS velocity estimates from the 16k simulations only, which has runs with the lowest initial string densities. Horizontal dashed lines and bands are as in Fig.~\ref{fig:xScaling}. There one can see that with the lowest string density the RMS velocity is fairly constant, at around $v\simeq 0.56$.  Here, as the initial string separation is comparable with the Hubble length, the Hubble damping is significant, and reduces the velocity in the initial burst of acceleration from rest.

\begin{figure}[h]
    \centering
    \includegraphics[width=\columnwidth]{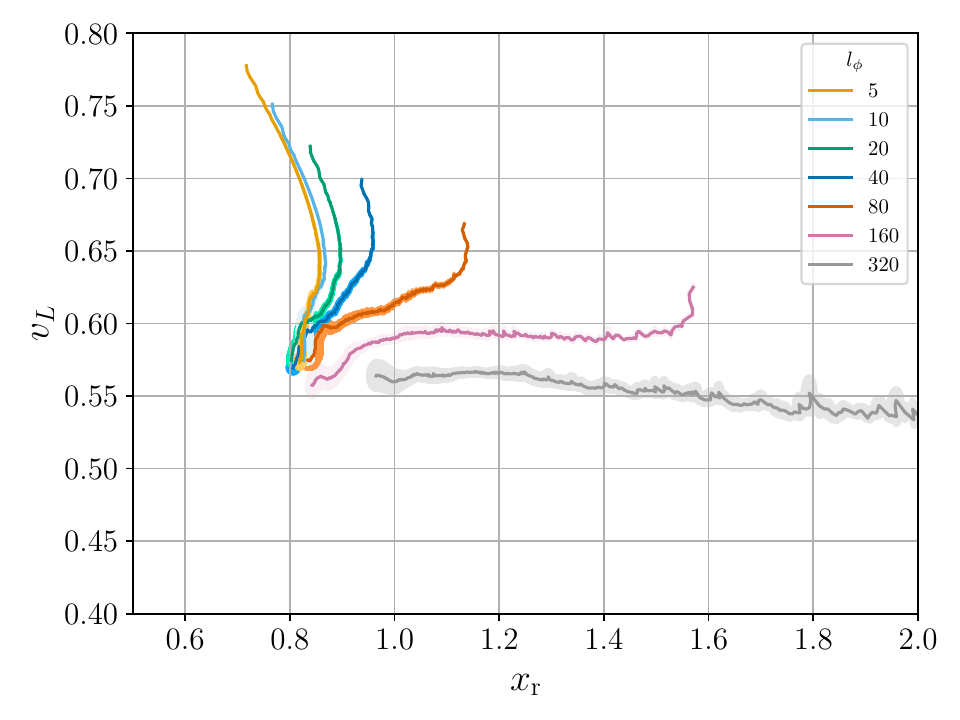}
    \caption{\label{fig:phase}
    Phase-space plot for the RMS velocity and $\xr$ (\ref{x})   for  the 16k simulations, and all initial correlation lengths $\IniCorLen$. 
    The semi-major and semi-minor axes of the shaded ellipses show 1$\si$ errors on the mean of the data points.
    }
 \end{figure}

The RMS velocity and the scaled mean string separation data can be combined into a phase-space plot (see Fig.~\ref{fig:phase}). Here we also include only the 16k simulations, with all initial field correlation lengths 

The visual impression of an evolution in the phase plane towards a point $(x,v) \simeq (0.8, 0.6)$ is reinforced. There are very few intersections of the averaged trajectories, which supports the idea that the dynamical evolution of these quantities is described by an autonomous system, as assumed by the VOS models described in Section \ref{sec:scaling}.  
The trajectories suggest that there is a roughly horizontal $v$ nullcline, with a segment in the area $0.55 \lesssim v \lesssim 0.60$ and $0.8 \lesssim x$, intersecting with a roughly vertical $x$ nullcline segment at $x \simeq 0.8$ in the same range of velocities.  This intersection would account for the fixed point at $(x,v) \simeq (0.8,0.6)$.

Outside that region the points where $dx/dv = 0$ suggest that there is also a roughly horizontal $x$ nullcline segment at $v \simeq 0.65$ in the range $0.8 \lesssim x \lesssim 1.2$. If or how this joins to the other nullcline segment is not clear.

\begin{figure}[t]
    \centering
    \includegraphics[width=\columnwidth]{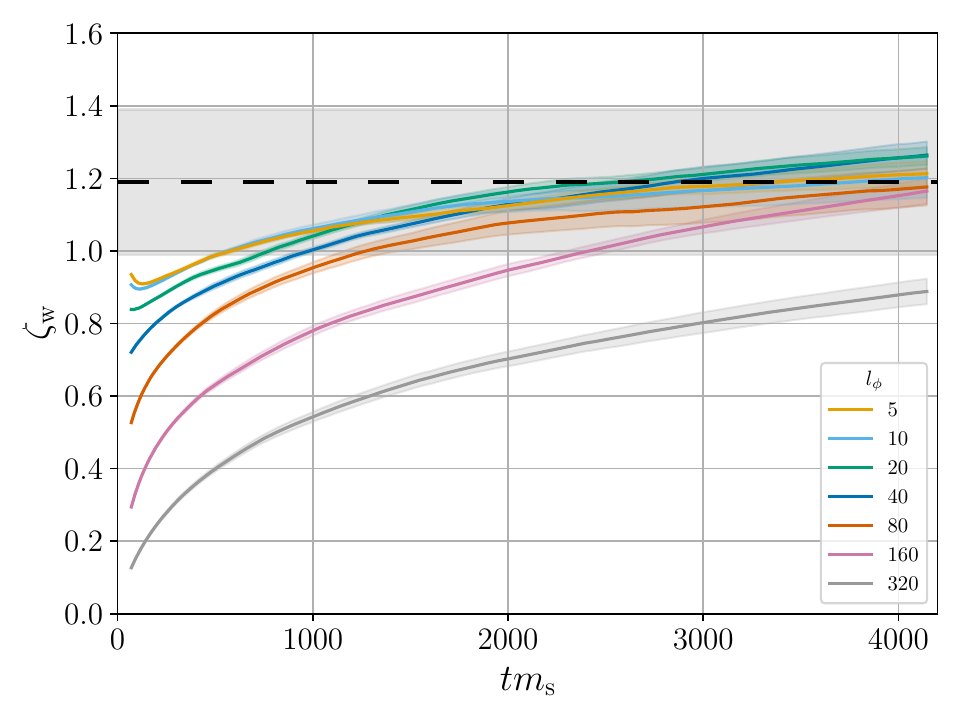}
     \includegraphics[width=\columnwidth]{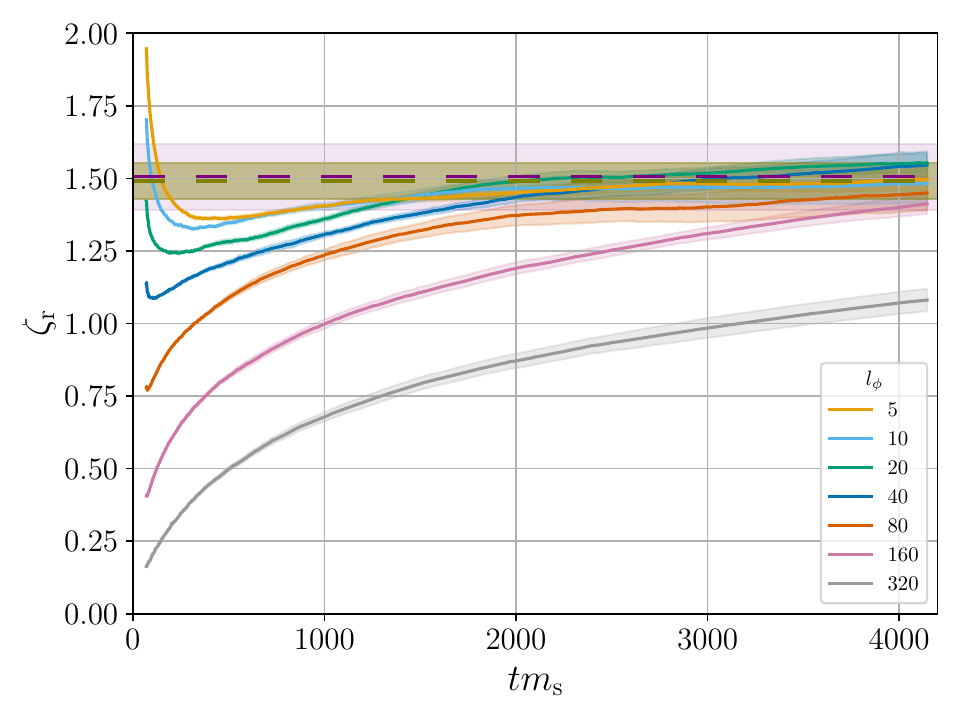}

    \caption{\label{fig:zeta} Plots of $\zeta_{\rm w}$ and $\zeta_{\rm r}$ against time for 16k simulations with the same colour coding as Fig. \ref{fig:xiScaling}. The horizontal dashed lines and shaded bands correspond, respectively, to the asymptotic mean values and $1\si$ errors obtained in \cite{Hindmarsh:2019csc} (top, in black), \cite{Hindmarsh:2021vih} (bottom, in purple) and this work (bottom, in olive green).}
 \end{figure}

\begin{figure}[ht]
    \centering
    \includegraphics[width=\columnwidth]{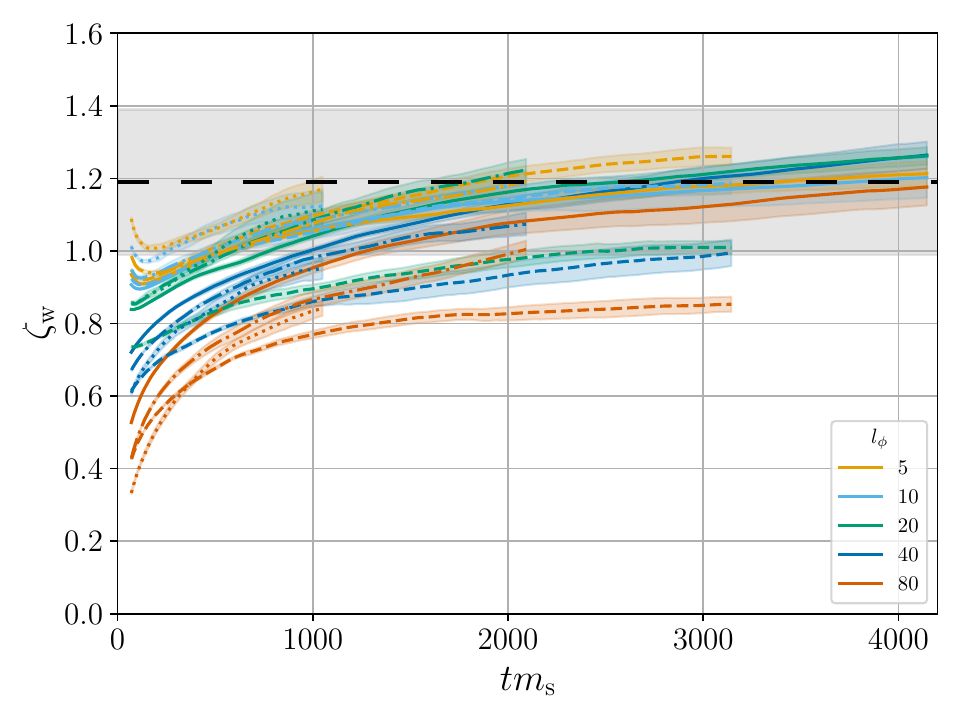}
    \includegraphics[width=\columnwidth]{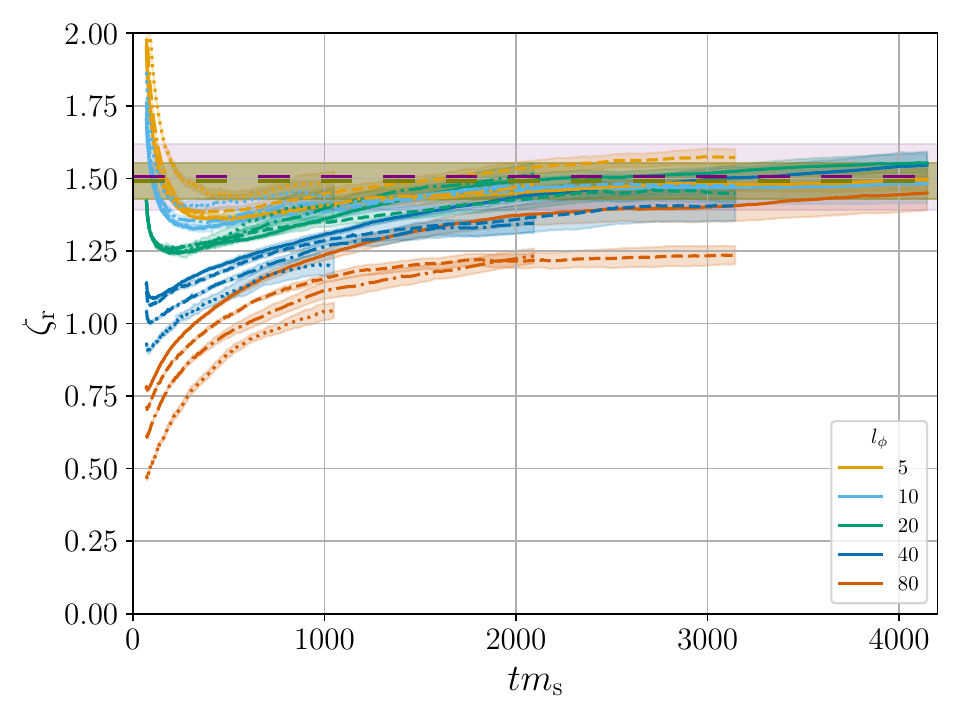}
    \caption{\label{fig:zetawindall} Plots of $\zeta_{\rm w}$ and $\zeta_{\rm r}$ against time for simulations for $\IniCorLen=(5,\ 10,\ 20,\ 40,\ 80)$ and all simulation box sizes. The shaded regions show the $1\sigma$ errors on the mean.  The horizontal dashed lines and shaded bands are as in Fig.~\ref{fig:zeta}.}
 \end{figure}

Returning to the string length, 
another choice to show its behaviour is to use the string length density scaled with cosmic time, $\zeta$  (\ref{e:zetaDef}).
The string length may be defined in the universe frame or in the string rest frame.  
 In Fig.~\ref{fig:zeta} we plot $\zew$ (top) and  $\zer$ (bottom) 
 for the 16k simulations, with all the different $\IniCorLen$, against cosmic time. 
 In  Figure \ref{fig:zetawindall} we only show the simulations for $\IniCorLen=5,10,20,40,80$, but for all sizes, i.e., 4k, 8k, 12k and 16k.  
The solid line is the average over realisations and the shaded bands are the 1$\sigma$ errors on the mean. 
The horizontal dashed lines and error bands show previous results for $\zew$ \cite{Hindmarsh:2019csc}, 
 $\zer$ \cite{Hindmarsh:2021vih}, and our new result for $\zer$, which we discuss below.

It can be seen that for simulations with lower $\IniCorLen$ (5,10,20), 
where the initial length density parameter (see Table \ref{t:ZetIni}) is greater than or approximately equal to $1.5$, 
$\zer$ appears to converge to a constant value for the last half of the simulation in cosmic time, whereas for higher values of $\IniCorLen$ (160, 320), 
where the initial length density parameter is much less than $1.5$, 
$\zer$ is still growing. We interpret this as the higher $\IniCorLen$ cases not having reached scaling yet. Note also that  the growing lines have not crossed the asymptotic value $\zeta_*$. 

The convergence is not so apparent when plotting $\zew$. The RMS velocity decreases slowly throughout the simulation for the simulations with lower $\IniCorLen$, which has the effect of reducing the Lorentz contraction, and hence increasing the universe-frame length relative to the string rest-frame length. 
However, in no case does the mean value of $\zew$ go above the 1$\si$ error band from previous simulations, and the largest value of $\zew$ over all simulations for $t\ms > 2000$ is $1.46$.

\subsection{Linear fit analysis}

In this subsection we repeat the analysis of \cite{Hindmarsh:2019csc} for a consistency check with previous results.
The agreement of the final values of $\zeta$ of the 16k simulations with the extrapolated value from the 4k data in \cite{Hindmarsh:2019csc} already supports the extrapolation methods used there to extract the asymptotic string density.

 \begin{figure}[h]
    \centering
    \includegraphics[width=\columnwidth]{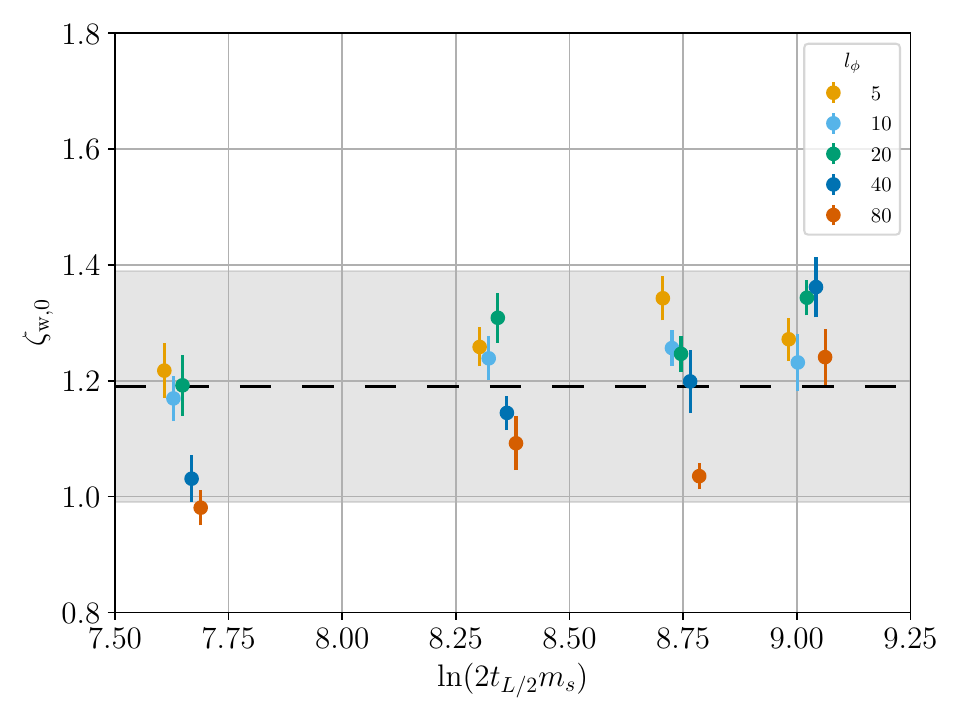}
     \includegraphics[width=\columnwidth]{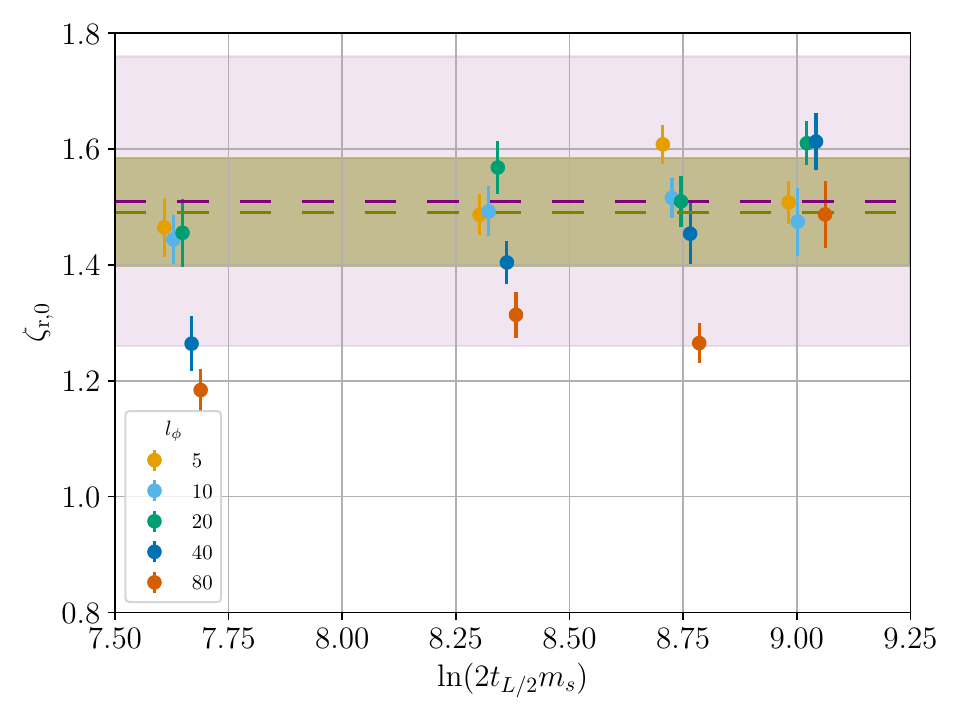}

    \caption{\label{fig:zetasize} Estimates of asymptotic values of length density parameters $\zew$ (top) and $\zer$ (bottom) obtained by linear fitting of mean string separations $\xiw$ and $\xir$ to cosmic time $t$, for all simulation box sizes (4k, 8k, 12k and 16k) and initial field correlation lengths $\IniCorLen$. On the $x$ axis is the logarithm of cosmic time at which light has travelled half way across the simulation box $\tHalf$, normalised by the twice scalar mass $\ms$. Small horizontal displacements have been applied in order to make data more visible. The horizontal dashed lines and shaded bands are as in Fig.~\ref{fig:zeta}.}

 \end{figure}

The extrapolation method of Ref.~\cite{Hindmarsh:2019csc} seeks an estimate of the asymptotic scaled string separation $x_*$ and the asymptotic length density $\ze_*$ from the slope of $\xi$ against $t$ towards the end of the simulation, for either of the two length measures $\xiw$ and $\xir$.  The details of the approach to scaling are accounted for by the intercept parameter of a linear least squares fit. 
This is a phenomenological tool to extract the slope, and not a prediction of how the mean string separation approaches scaling.

The fit is carried out on the physical mean separation against cosmic time, 
from which a estimates of the slopes $\xrLinEst$, $\xwLinEst$ for both of the length measures are obtained, as outlined below. 
The asymptotic scaled string separations are just the estimates of the slopes, and the asymptotic length density estimates are 
\ben
\zerLinEst = 1/\xrLinEst^2, \quad
\zewLinEst = 1/\xwLinEst^2.
\label{e:zetabeta}
\een 
We obtain the slope estimates by least squares fits  to 
\ben
\xi = \FitCon_b + \FitSlo_b t\,,
\label{eq:linfit}
\een
over time intervals $t_b < t < t_{b+1}$, with $0 \le b \le N_b$.  Specifically, $N_b = 4$, and the interval boundaries are chosen at conformal times $[6,7,8,9,10] \tau_{s}$, where $\tau_s = 100 n/4$ for the $n$k simulations. Hence, for the 16k simulations, the largest boundary is $\tau = 4000$. These times are in the later part of the simulation, where the system has had the longest opportunity to lose memory of the initial conditions.  

The slopes are then averaged over runs (random seeds) to give an estimate 
$\FitSlo_b(\IniCorLen, n)$ in each time interval, initial correlation length and box size. 
The estimate of the asymptotic slope for a given initial field correlation length and box size is the mean of $\FitSlo_b(\IniCorLen, n)$, that is,
\ben
\xLinEst = \frac{1}{N_b} \sum_b \FitSlo_b(\IniCorLen, n).
\een

The uncertainties are estimated as follows. First, we compute the standard error on the mean of the slopes for a given interval, which we denote $s_b(\IniCorLen, n)$. This is always much greater than the square root of the fit covariances on the slope parameters. 
The uncertainty $u$ on $\hat\FitSlo$ is constructed from the 
RMS of the standard errors on the means and the standard deviation of the means, added in quadrature:
\ben
\hat{u}^2(\IniCorLen, n) = \sum_b \left( \frac{1}{N_b-1} (\FitSlo_b -\xLinEst)^2 + \frac{1}{N_b}s_b^2  \right) .
\een
Finally, 
we convert the slope estimate for the mean separation in each frame (universe frame, string rest frame) to estimates of the asymptotic length density parameter $\zewLinEst$, $\zerLinEst$ as per Eq.~\eqref{e:zetabeta}.
The results are plotted in Fig.~\ref{fig:zetasize}. Again, the dashed lines and error bands show previous results for $\zew$ \cite{Hindmarsh:2019csc},  $\zer$ \cite{Hindmarsh:2021vih}, and our new result for $\zer$.

We see that the consistency of the slopes between simulation sets depends on the initial field correlation length, and hence the string density at the start of the physical evolution.  The higher-density simulations $\IniCorLen = 5,10$ have length densities which are consistent between simulations with $N$ = 4k, 8k, 12k and 16k.  However, length density estimates for simulations with $\IniCorLen = 40, 80$ are low, but increasing with simulation size. At 16k the length density estimates from simulations with  $5 \le \IniCorLen \le 80$ are in agreement.

\subsection{Asymptotic length densities and RMS velocity}
\label{sec:asymLv}

\begin{figure}[htbp]
\begin{center}
\includegraphics[width=\columnwidth]{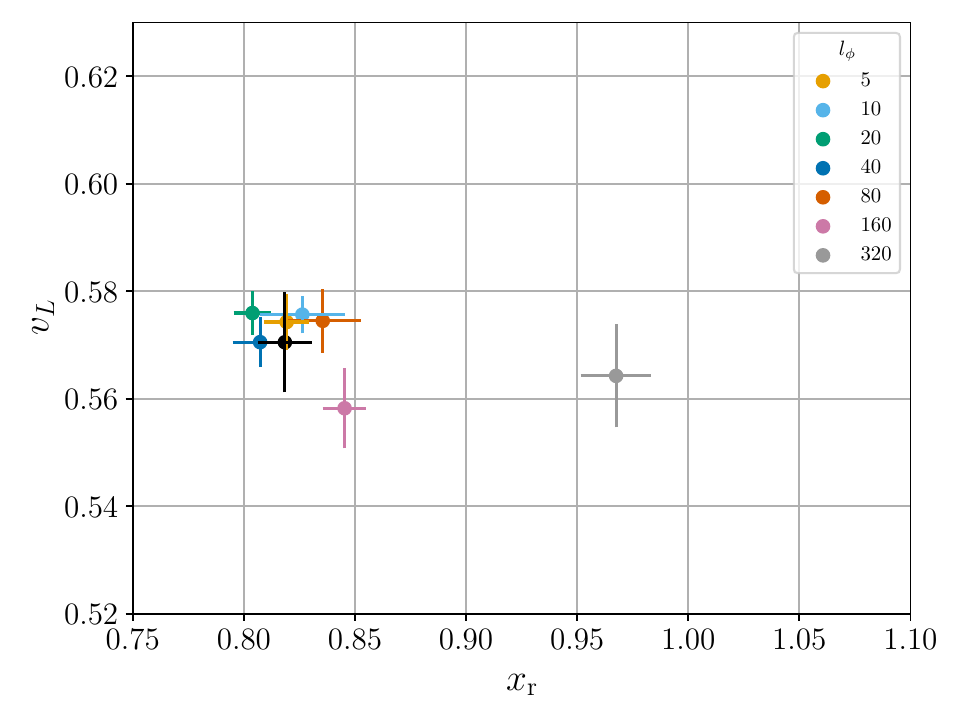}
\caption{
Mean and uncertainties of the scaled mean separation $\xr$ and the RMS velocity $v_{\rm s}$ of the 16k simulations at the nearest data point to conformal time $\tau = L/2$, where $L$ is the conformal length of the simulation box side. At this time $t\ms = 4044$.  
The black data point shows the mean over $\xr$ for $5 \le \IniCorLen \le 80$ and $v_L$ for all $\IniCorLen$.
The computation of the uncertainties is described in the text.
}
\label{fig:x_v_last_16k}
\end{center}
\end{figure}

In Table \ref{t:x_v_last_16k} we give the 
averages of $\xr$ and $v_L$  over 16k simulations at $\tau = L/2$, where $L$ is the conformal length of simulation box sides. This is the time at which light signals moving in opposite directions start to meet, and so a conservative upper bound on the times over which the periodic boundary conditions do not affect the evolution. 
The same results are plotted in Fig.~\ref{fig:x_v_last_16k}. 
One can see a statistical consistency in $\xr$ between simulations with  $5 \le \IniCorLen \le 160$, and in  $v_L$ between all simulations.  
The black point and cross-hairs show the combined $\xr$ data from $5 \le \IniCorLen \le 80$ and $v_L$ data from all simulations. 
We explain this choice below. 

\begin{table}
\begin{tabular}{l l l}
$l_\phi$ & {\hfill $\hat{x}_\text{r}$ \hfill} & {\hfill $\hat{v}_{L}$ \hfill} \\
\hline
5 & $ 0.822(10)$  & $ 0.5743(51)$ \\
10 & $ 0.829(19)$  & $ 0.5757(35)$ \\
20 & $ 0.8069(85)$  & $ 0.5760(40)$ \\
40 & $ 0.810(12)$  & $ 0.5706(46)$ \\
80 & $ 0.839(17)$  & $ 0.5745(59)$ \\
160 & $ 0.8485(99)$  & $ 0.5583(74)$ \\
320 & $ 0.971(16)$  & $ 0.5643(96)$ 
\end{tabular}
\caption{
Estimates of scaled mean string separation $\xr$ and RMS velocity $v_\text{s}$ for 16k simulations at the cosmic time $\tHalf = 4044/\ms$, close to the end. The computation of the uncertainties is described in the text. \label{t:x_v_last_16k}
}
\end{table}

\begin{figure}[h]
    \centering
     \includegraphics[width=\columnwidth]{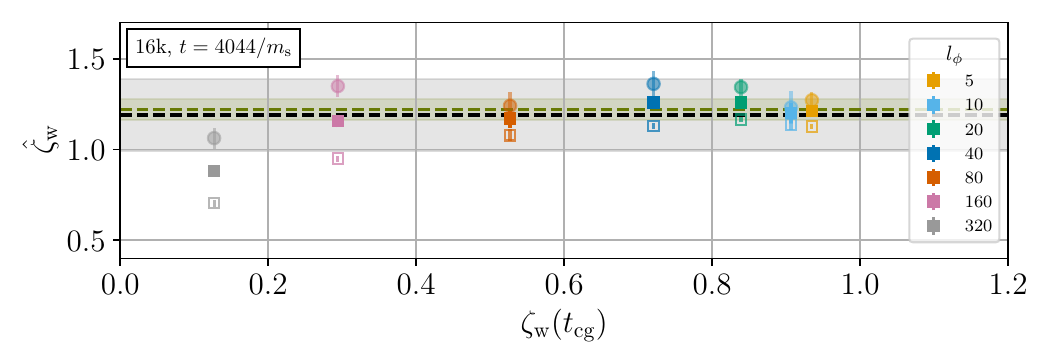}
     \includegraphics[width=\columnwidth]{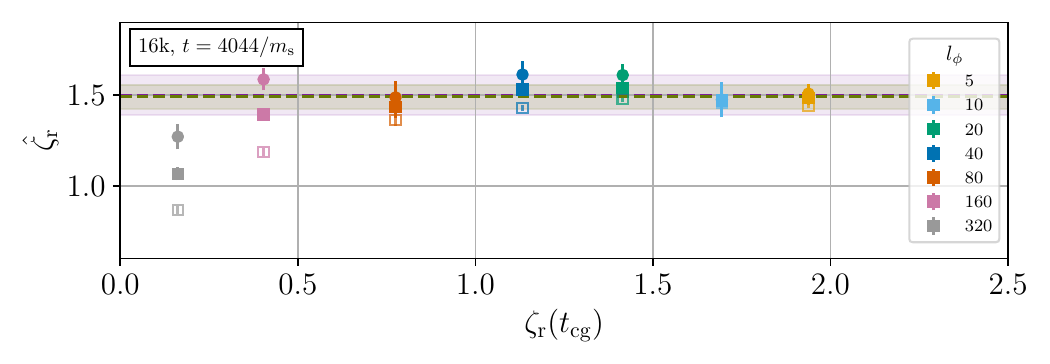}

    \caption{\label{fig:zetafinal} Estimates of asymptotic values of length density parameters $\zew$ (top) and $\zer$ (bottom), in the largest (16k) simulations. 
Estimates are obtained by linear fits (circles), at cosmic time $t_{L/2}$ (filled squares).
Also shown are the values at $t_{L/2}/2$ (empty squares). The olive green dashed line and band shows the mean and 1$\si$ uncertainty of runs with $5 \le \IniCorLen \le 80$. The grey dashed lines and bands (top) is the equivalent result from \cite{Hindmarsh:2019csc}, and the purple dashed lines and bands (bottom) is the equivalent result from \cite{Hindmarsh:2021vih}.  
}
 \end{figure}

To study the convergence of the results in the 16k simulations,
in Fig.~\ref{fig:zetafinal} we show both the fitted values of $\zeta$ (filled circles) with those at the nearest time to $t_{L/2}$ (filled squares). Also shown are the values at time nearest to $t_{L/2}/2$ (empty squares).  They are plotted against 
the length density parameter at the start of the physical evolution, $\zer(t_\text{cg})$.  Data from all box sizes is plotted in Appendix \ref{s:ConAll}.
 
The length density at $t_{L/2}$ is systematically lower  in the 16k simulations for simulations with $\zer(t_\text{cg})  \lesssim 0.5$ ($160 \le \IniCorLen \le 320$).  For 16k simulations with $\zer(t_\text{cg})  \gtrsim 0.5$ ($5 \le \IniCorLen \le 80$), the agreement between the linear fits and the value at $t_{L/2}$ is very good.  For $\IniCorLen = 5,10$, the values $\zer(t_{L/2})$ and $\zer(0.5\tHalf)$ are in agreement.  

We therefore consider it justified to take our final estimate of the asymptotic length density parameter to be the mean of the final values in the 16k simulations with $5 \le \IniCorLen \le 80$.
The uncertainty is computed as the sum in quadrature of the RMS errors on the mean and the standard deviation of the means.
The result is
\ben
\hat\zeta_{\text{r},*} = \zzRestMeanAxCESS(\zzRestErrsAxCESS) .
\label{e:zetaRestResult}
\een
This result is shown as an olive green dashed line and band in Fig.~\ref{fig:zetafinal}.
This is consistent with our previously published estimate $\hat\zeta_{\text{r},*} = 1.50(11)$ \cite{Hindmarsh:2021vih}.
The corresponding values for the scale mean string separation is 
\ben
\hat{x}_{\text{r},*} = \xRestMeanAxCESS(\xRestErrsAxCESS) .
\label{e:xRestResult}
\een

\begin{figure}[h]
    \centering
     \includegraphics[width=\columnwidth]{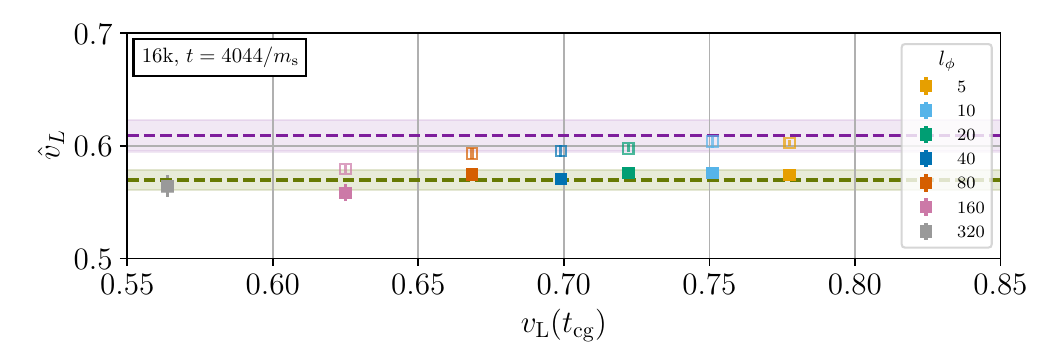}

\caption{\label{fig:vfinal}  Estimates of asymptotic values of RMS velocity $v_L$, in the largest (16k) simulations, obtained at cosmic time $t_{L/2}$ (filled squares). 
Also shown are the values at $t_{L/2}/2$ (empty squares). The olive green dashed line and band shows the mean and 1$\si$ uncertainty of all runs. The purple dashed lines and bands is the equivalent result from \cite{Hindmarsh:2021vih}.
 }
 \end{figure}

In Fig.~\ref{fig:vfinal} we show the RMS velocities $v_L$ at $t_{L/2}$ and $t_{L/2}/2$ for 16k runs, plotted against the velocities at the start of physical evolution $v_L(\tcg)$. The colour scheme and markers follow the same conventions as in the previous figures. 
The purple dashed line and band is the equivalent result from \cite{Hindmarsh:2021vih}.

We see that there is evolution of the velocity by about 5\% between the two times plotted, except for the largest initial field correlation length $\IniCorLen = 320$, which stays consistently at around $v_L \simeq 0.57$. By the half light-crossing time $\tHalf$, all velocities are consistent.  
We take the agreement of the runs with $\IniCorLen = 320$ as a justification for combining all the data as our 
estimate for the asymptotic RMS velocity 
\ben
\hat{v}_{L,*} = \vLMeanAxCESS(\vLErrsAxCESS) .
\label{e:vRestResult}
\een
The uncertainty computed in the same way as for $\zer$. 
This result is shown as an olive green dashed line and band in Fig.~\ref{fig:vfinal}. Note that it is statistically inconsistent with our previous result using the scalar velocity estimator and fitting to a two-parameter VOS model $\vs = \vsStarVOS(\dvsStarVOS)$.

\begin{table}[htp]
\begin{center}
\begin{tabular}{l | c | c }
$\IniCorLen$ & $\hat{\ze}_\text{w}$ & $\hat{\ze}_\text{r} \gamma^{-1}(\hat{v}_L)$ \\
\hline
5 & $ 1.211(34)$  & $ 1.214(24)$ \\
10 & $ 1.200(56)$  & $ 1.202(46)$ \\
20 & $ 1.258(24)$  & $ 1.259(21)$ \\
40 & $ 1.259(37)$  & $ 1.257(30)$ \\
80 & $ 1.172(51)$  & $ 1.174(42)$ \\
160 & $ 1.156(34)$  & $ 1.156(24)$ \\
320 & $ 0.882(32)$  & $ 0.880(34)$ 
\end{tabular}
\end{center}
\caption{\label{t:zetawest} Estimates of universe-frame length density parameter for 16k simulations, for initial field correlation lengths $\IniCorLen$.  In the second column are the directly measured values at cosmic time $t_{L/2}$ (conformal time $\tau = L/2$, and in the third the values inferred using Eq.~\eqref{rela}.}
\end{table}%

The continuing evolution of the velocity from the high values created by the initial conditions gives an explanation for the slowness of the convergence of $\zew$, which is related to the rest-frame length density parameter and the RMS velocity through Eq.~\eqref{rela}.
Hence, as long as the velocity still displays some evolution in the analysed range, the Lorentz contracted (universe frame) version of the string length parameter $\zew$ will also evolve.
Conversely, the simulations for which the velocity is approximately constant ($\IniCorLen = 320$) start at low density, and there is strong evolution in $\zer$ instead.  The combination $\zer \ga^{-1}(v_L)$ therefore evolves for all simulations.  

In Table \ref{t:zetawest} we give the values of $\zew$ at $t_{L/2}$ calculated both directly and through \eqref{rela} for the 16k simulations.  The agreement is extremely good. If our argument that the RMS velocities have converged is correct, then this agreement signals that $\zew$ has also converged where $\zer$ has.  Its mean and uncertainty, calculated in the same way as for $\zer$, is
\ben
\zewAsymEst = \zzWindMeanAxCESS(\zzWindErrsAxCESS).
\een

\section{Long-term growth analysis}
\label{s:LonTerGro}

In this section, we analyse our results in the context of the logarithmic growth model \cite{Gorghetto:2020qws,Gorghetto:2021fsn}, which starts from the observation that 
at low string densities,  the universe-frame length density parameter $\ze_{\text{w}}$ is observed to grow rather slowly, and approximately proportionally to the logarithm of cosmic time, with a proportionality constant $c_1 \simeq 0.2$. 

 \begin{figure}[h]
    \centering
         \includegraphics[width=\columnwidth]{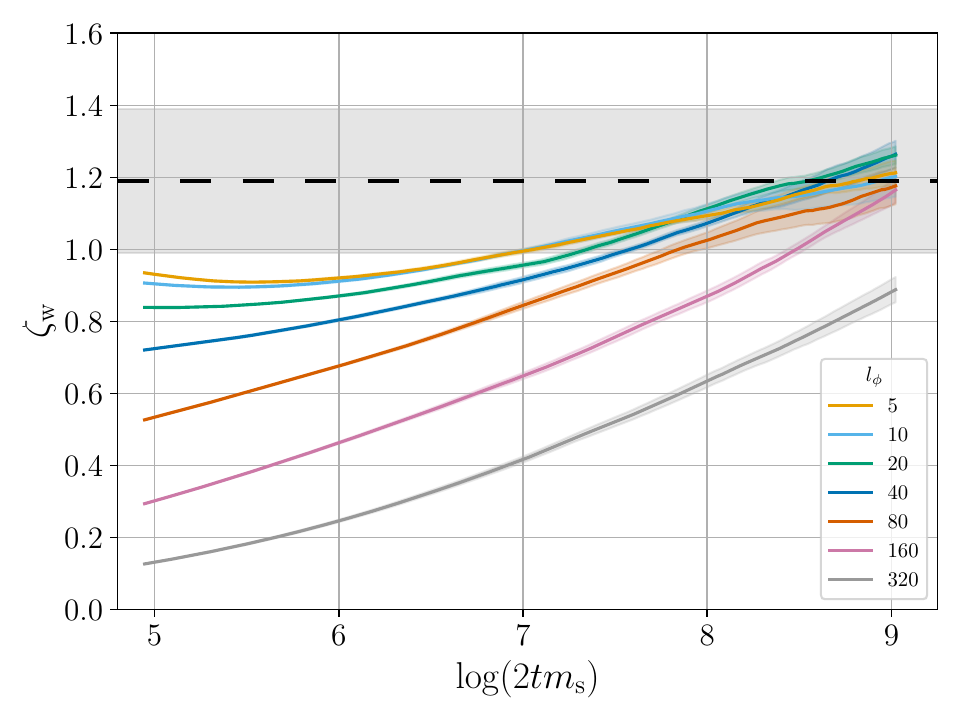}
        \includegraphics[width=\columnwidth]{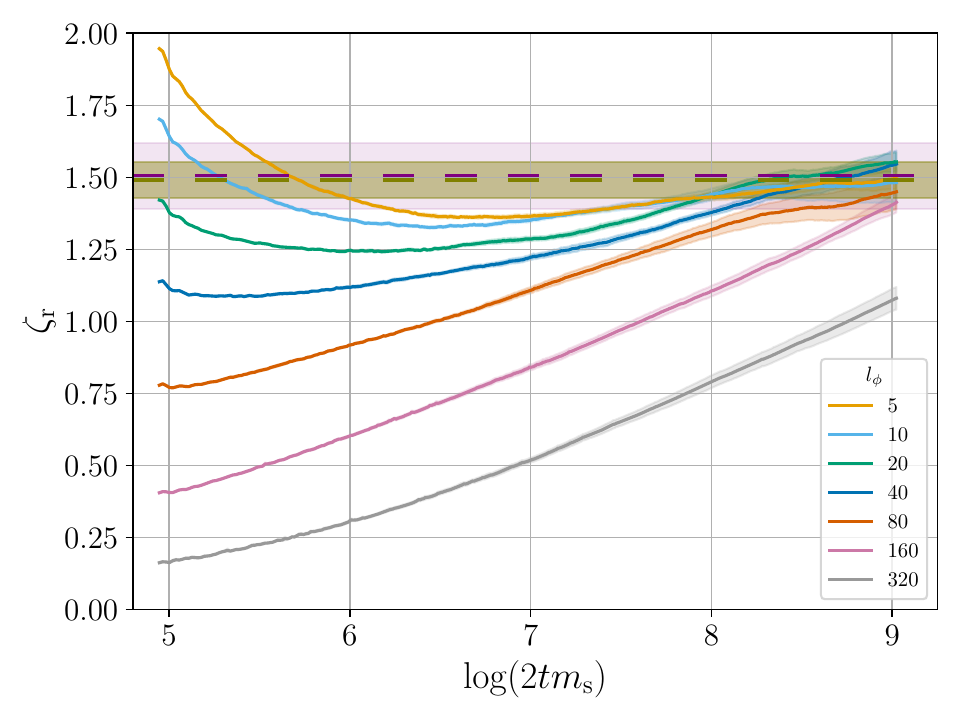}
    \caption{\label{fig:ZetLog} Plots of $\zeta_{\rm w}$ and $\zeta_{\rm r}$ against logarithm of time for 16k simulations . This figure is analogous to Fig.~\ref{fig:zeta}, but in logarithmic scale.  }
 \end{figure}
 
  \begin{figure}[h]
    \centering
    \includegraphics[width=\columnwidth]{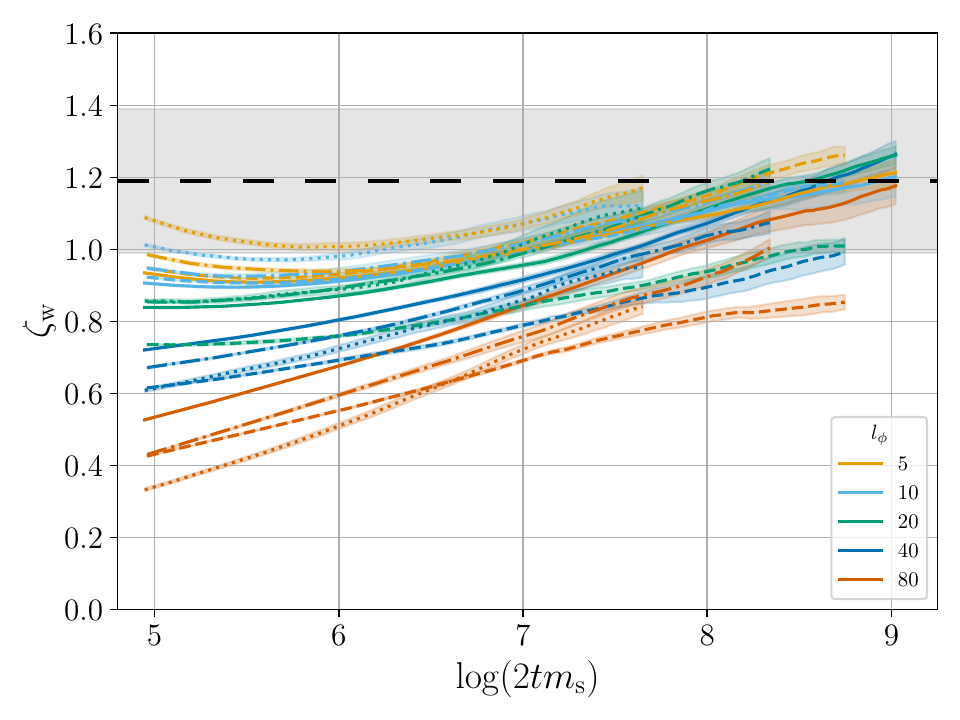}
    \includegraphics[width=\columnwidth]{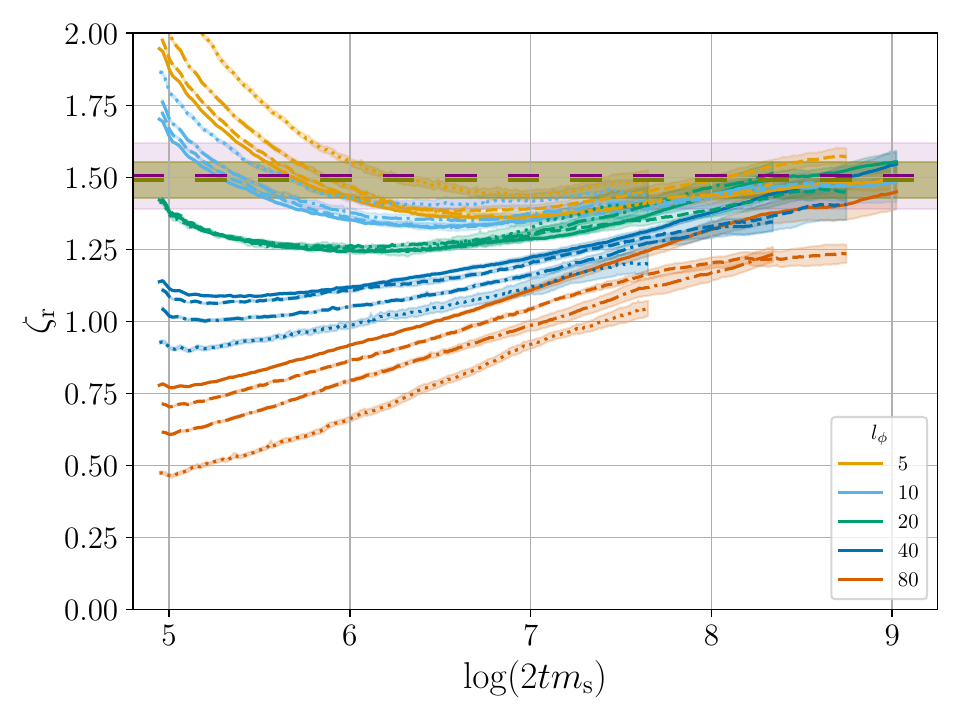}
    \caption{\label{fig:zetawindall_log} Plots of $\zeta_{\rm w}$ and $\zeta_{\rm r}$ against the logarithm of time for simulations for $\IniCorLen=5,\ 10,\ 20,\ 40,\ 80$ and all simulation box sizes. This figure is analogous to Fig.~\ref{fig:zetawindall} but in logarithmic scale.}
 \end{figure}

First, we plot the universe frame and rest frame length density parameters $\zew$ and $\zer$ against  $\log(2t\ms)$, equal to $\log(\ms/H)$ in a radiation-dominated universe, to facilitate comparison with other groups' results. 
This choice of variable is the natural one to study evolution of the dynamical system towards fixed points, although it tends to accentuate the early times, where the effect of the initial conditions of the simulation is stronger. It also compresses the latest times of the simulation, where in principle the simulations are less sensitive to the initial conditions.

In Fig.~\ref{fig:ZetLog} we show results from the 16k simulations for all initial field correlation lengths, and in Fig.~\ref{fig:zetawindall_log} the results from all box sizes (4k, 8k, 12k and 16k) for the common initial field correlation lengths ($\IniCorLen	 = 5, 10, 20, 40, 80$). 
Simulations with different box sizes start with different RMS velocities, and hence evolve differently.

To test the logarithmic growth model, we fit the form
\ben
\ze_{\text{w}} = c_0 + c_1\ln(2t m_{\rm s})
\label{eq:logfit}
\een
to our data from the largest (16k) simulations over ranges $\log(2 t \ms) \in [6,7]$, $[7,8]$, and $[8,9]$. We thereby obtain least-squares estimates of the slopes $c_1$ at different times, and hence different length densities. These estimates were then averaged over multiple runs. 

\begin{figure}[ht]
    \centering
    \includegraphics[width=\columnwidth]{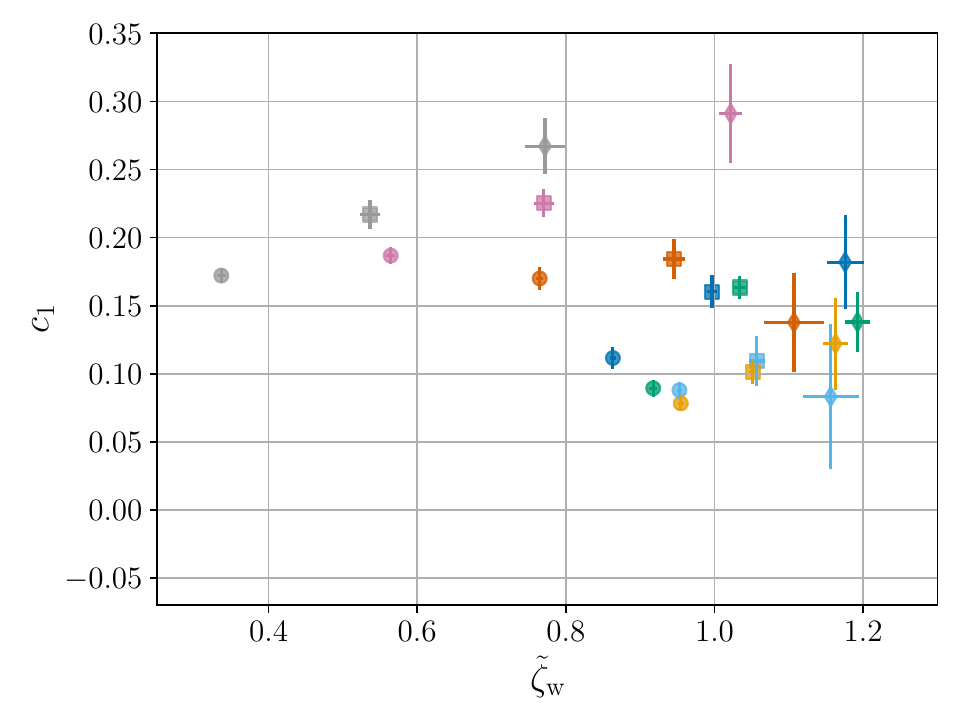}
    \includegraphics[width=\columnwidth]{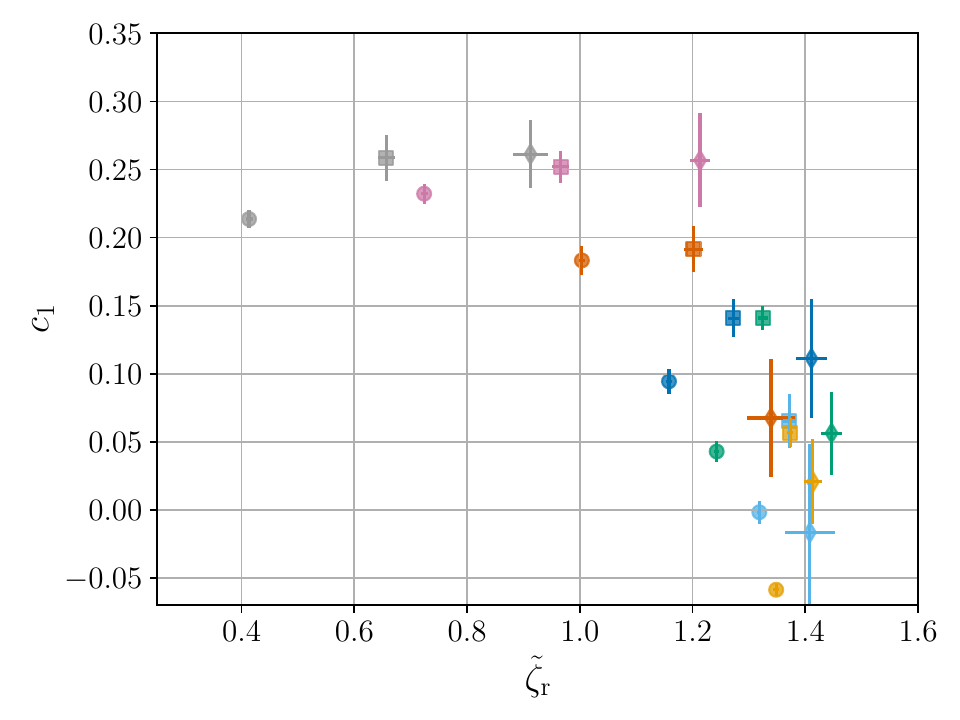}
    \caption{\label{fig:logslopes} Parameters from logarithmic fit (\ref{eq:logfit}) to $\zeta_{\rm w}$ (upper panel) and $\zeta_{\rm r}$ (lower panel) versus the median value of $\zeta$ in the ranges $\ln(2t m_{\rm s})=[6,\ 7],\ [7,\ 8],\ [8,\ 9]$ represented with circles, squares and diamonds respectively for all correlation length. The color coding corresponds to that of the 16k simulations (see for example Fig.~\ref{fig:ZetLog}).}
 \end{figure}

In Fig.~\ref{fig:logslopes} we plot the slopes $c_1$ against the median length density parameters over the fit ranges, $\tilde\ze_{\rm w}$ and $\tilde\ze_{\rm r}$. If the logarithmic growth persists, one should see a convergence towards a consistent slope at large time, and hence large $\ze$. 

Considering first the universe-frame length densities, 
for $\zew \lesssim 0.8$, the fitted slopes of $\zew$ are approximately $c_1 \simeq 0.2$, consistent with other groups' observations. The two sets of simulations with the lowest initial length density show a tendency towards a steepening of the slope with $\tilde\ze_{\rm w}$ and hence time.
For larger values of $\tilde\ze_{\rm w}$, which are found towards the end of the simulation, the slopes are somewhat smaller.  For the two sets of simulations with the highest initial string density, the slopes are consistently around $c_1 \simeq 0.1$. 

Turning to the rest-frame length density parameter $\ze_{\rm r}$, the slopes for $\tilde\ze_{\rm r} \lesssim 1.0$ are a little steeper, at around $c_1 \simeq 0.25$.  For larger $\tilde\ze_{\rm r}$ 
there is a clear tendency for the slope to get smaller. For $\zer$, the slopes of the two sets of simulations with the highest initial string density are consistent with zero in the final time interval. 
The difference in slopes between $\zew$ and $\zer$ is due to the evolution of the RMS velocity.

In the logarithmic growth picture, the lower values of $c_1$ at higher $\ze$ are considered to be transients which will eventually die out as the assumed long-term logarithmic growth in $\ze$ emerges.  The slopes should increase with time and hence $\ze$. 
This steepening is not manifest in $\zew$, although the uncertainties in the slope in the final interval are large. 
 In $\zer$, consistent steepening is visible only for the lowest string density .

Other groups tend to favour low-$\ze$ initial conditions with $\zew(5) \simeq 0.5$, and reach $\zew \simeq 1.2$ at around $\log(2t\ms) \simeq 9$ (see Table \ref{t:others}). This is close to the behaviour of our 16k simulations with $\IniCorLen = 80$.  
As shown in the last section, our $\IniCorLen = 80$ simulations are converged by the agreement of the estimate of $\zer$ derived from linear slope of the mean string separation $\xir$ with its value at $\tHalf$.  The logarithmic slope over the final interval (corresponding to $ 1480 \le t \ms  \le 4052$) is $c_1 = 0.07(4)$, 
less than $2\si$ away from zero, which also indicates a system close to convergence over that interval. 
In the same set of simulations, $\zew$ has a steeper slope, around $c_1 = 0.13(3)$, due to the velocity evolution. Our study of the  velocity in the previous section indicates that the velocity has converged for all simulations by $\log(2t\ms) = 9$.  If this is correct, we would expect to see the slopes of the $\zew$ sharing the evolution towards zero for still higher values of  $\log(2t\ms)$.

\begin{table}
\begin{tabular}{l | c | c | l}
Publication &  $\zew(5)$ & $\max(\La)$ & {\hfill $\max(\zew)$  \hfill}\\
\hline
Klaer, Moore (2019) \cite{Klaer:2019fxc}, Fig.~2 & $0.59$ & 7.3 & 0.96(6)  \\
Gorghetto et al. (2021) \cite{Gorghetto:2021fsn}, Fig.~1 & $0.50$ & $7.9$ &  $0.99(6)$ \\
Buschmann et al.(2021) \cite{Buschmann:2021sdq}, Fig.~2 & $0.42$ & $9.0$ & $1.22(16)$ \\
Saikawa et al. (2024) \cite{Saikawa:2024bta}, Fig.~7  & $0.50$ & $9.1$ & 1.20(3) \\
Kim et al (2024) \cite{Kim:2024wku}, Fig.~3 & $0.45$ & $7.9$ & 0.90(2)
\end{tabular}
\caption{\label{t:others} 
Summary of recent universe-frame length density data from other groups. Given is: mean length density at $\La \equiv \log(2 t \ms) = 5$, the maximum value of $\La$ reached, and the value of $\zew$ at that time, with uncertainty.  All values estimated digitally \cite{WebPlotDigitizer} from the quoted figures.
}
\end{table}

An exception to the tendency to measure $\zew$ starting from low densities can be found in Ref.~\cite{Klaer:2017qhr}, where $\zer$ from two sets of simulations with initial values approximately 1 and 2 are shown (Fig.~3).  In these simulations, the logarithmic slope remains small, and final values of the rest-frame length density are both $\zer \simeq 1.4\pm 0.1$, consistent with our results for the fixed point. 

\section{Discussion and conclusions}
\label{sec:Con}

In this paper we have studied the evolution of the global properties of axion string networks, the $t$-scaled length densities 
$\zer$ (from length measured in the string rest frame), $\zew$ (from length measured in the cosmic rest frame),  and the RMS velocity of the strings $v_L$ (estimated from the Lagrangian density of the fields in the string core).  We have carried out a programme of simulations between 4k and 16k grid points per side, the largest fixed grid simulations to date. The largest of our simulations have a dynamic range of a factor $60$ in cosmic time, starting from the cosmic time at the start of physical evolution $t_\text{cg} = 70/\ms$, where $\ms$ is the scalar mass, and therefore the scale of the width of the string core.  They reach a mean string separation, estimated from the string length in the universe frame, of $\xiw \simeq 3900/\ms$.

In order to investigate the effect of initial conditions, we start from a range of initial length densities differing by a factor 20, and a range of initial RMS velocities between $0.5$ and $0.8$. In all cases, by the end of the simulations, $1.0 \lesssim \zer \lesssim 1.5  $ and $0.56  \lesssim \vs \lesssim 0.61$.
They are consistent with the existence of a fixed point with values given in Table \ref{t:SumRes}.
Simulations which start out with length densities close to the fixed point value maintain a length density close to the fixed point value, while those which start out with lowest initial length densities evolve towards the fixed point but do not reach it. Similarly, simulations whose velocity starts out close to the fixed point value stay close.  
These tend to have strong length density evolution in our method of generating initial conditions.

Because of the velocity evolution, the universe-frame string length density parameter $\zew$ evolves more strongly than the rest-frame length density $\zer$. However the values at the end agree with the Lorentz-contracted rest-frame length density, which suggests that the universe length density is close to converging too. 

We also showed the evolution of the system in terms of motion through the phase plane $(\xr, v_L)$, which gives a strong visual impression of a fixed point around $(0.8, 0.6)$. Such evolution should be understandable in terms of 
a VOS model, which describes the average dynamics of the string network as a dynamical system with the logarithm of time as the evolution variable. 
Our larger simulations 
do not appear to follow a spiral approach close to the fixed point as predicted by the two-parameter VOS model of \cite{Martins:2000cs,Hindmarsh:2021vih}, and our new value for the fixed point velocity is about $5\%$ lower than the model predicts, using 4k simulations \cite{Hindmarsh:2021vih}.
Nevertheless, the phase space evolution does seem to be compatible with there being nullclines for $\xr$ and $v_L$, whose intersection creates a fixed point. A more thorough investigation of the phase plane, supporting the development of an improved VOS model is needed. Here, the work of Ref.~\cite{Klaer:2019fxc} already contains useful information.

We also analysed the evolution of the length density parameters looking for evidence of consistent long-term growth, 
by studying the slope of $\zew$ and $\zer$ against $\log(2t\ms)$.  If a logarithmic growth \cite{Gorghetto:2019hee,Gorghetto:2020qws}
is the long-term behaviour, we would expect to see a consistent slope emerging at late times and higher length densities.

For low densities, $\ze \simeq 0.5$, we find that the slope of  $\zew$  is around $0.2$, consistent with other groups, but no clear value was obtained.  The slope tends to be smaller for densities closer to the scaling fixed point values ($\zew \simeq 1.2$, $\zer \simeq 1.5$).  This tendency is especially clear for the rest-frame measure, where there is a cluster of slopes around zero, consistent with an approach to a fixed point. 

Other groups \cite{Klaer:2019fxc,Saikawa:2014pgv} have proposed that the system is chasing a quasi-fixed point with a logarithmic scale-dependence, which gives rise to a more complicated growth function.  Its form depends on the assumed response of the length density to displacements from the moving target. While a good fit can be found for a particular initial string density \cite{Saikawa:2014pgv}, the fit relies on low-density data. The agreement in our data of the value of the scaling fixed point obtained from box size 4k simulations \cite{Hindmarsh:2021vih} 
with that obtained in the new simulations at box size 16k implies that the evolution of the fixed point is small, and consistent with zero.

\begin{table}[htp]
\begin{center}
\begin{tabular}{c|l}
Quantity & {\hfill Value \hfill} \\
\hline
$\zewAsymEst$	& $\zzWindMeanAxCESS(\zzWindErrsAxCESS)$	\\
$\zerAsymEst$	& $\zzRestMeanAxCESS(\zzRestErrsAxCESS)$	\\
$\xrAsymEst$ & $\xRestMeanAxCESS(\xRestErrsAxCESS)$ \\
$\vLAsymEst$	&	$\vLMeanAxCESS(\vLErrsAxCESS)$
\end{tabular}
\end{center}
\caption{\label{t:SumRes}
Summary table of results for the scaling fixed point of an axion string network in the simplest field theory model studied in this paper, with field equation \eqref{eq:eom}. See Section \ref{sec:ModelSims} for definitions, and Section \ref{s:ResStaScaAna} for the analysis.}
\end{table}

In conclusion, our results support standard scaling of axion string networks with a well-determined fixed point in string density and RMS velocity, which we summarise in Table \ref{t:SumRes}. 
The observation of slowly increasing $\zew$, is a transient associated with low string density, and slow velocity evolution.

The accuracy of the results would be improved by better control of the initial conditions, so that the system starts out closer to the fixed point, and can approach it more closely.  While our method of preparation of initial conditions creates networks which are close to the length density fixed point, or the velocity fixed point, it does not manage both at the same time. 
For the future, it seems that all groups would benefit from developing and using such a method.

\begin{acknowledgments}
We are grateful to Daniel Cutting and Eelis Mielonen for collaboration on the initial stages of the project. We are grateful to the organisers of the workshop "IBS CTPU-CGA 2023 Workshop on Topological Defects" held at the Institute of Basic Sciences in Daejeon and the organisers of the follow-up workshop "Topological defects in Cosmology" at the University of Manchester. JC is grateful for discussions at these workshops with participants Amelia Drew, Christophe Ringeval, Tasos Avgoustdis, Richard Battye, Paul Shellard, Carlos Martins, Ivan Rybak and Masahide Yamaguchi. 
MH (ORCID ID 0000-0002-9307-437X) acknowledges support from the 
Research Council of Finland (grant number 333609). JL (ORCID ID 0000-0002-1198-3191), ALE (ORCID ID 0000-0002-1696-3579) and JU (ORCID ID 0000-0002-4221-2859) acknowledge support from Eusko Jaurlaritza IT1628-22 and by the PID2021-123703NB-C21 grant funded by MCIN/AEI/10.13039/501100011033/ and by ERDF; ``A way of making Europe”. JC (ORCID ID 0000-0002-3375-0997) acknowledges support from Research Council Finland grant 354572. 
KR (ORCID ID ID 0000-0003-2266-4716) acknowledges support from the Research Council of Finland grants 345070, 320123 and
319066, and the European Research Council grant 101142449.
Our simulations made use of Lumi at CSC Finland under pilot access project AxCESS with 1.1 M GPU-hours.  
\end{acknowledgments}

\appendix

\section{Determination of string solution parameters  $\mu_{V}$ and $f_{V,V}$}
\label{a:NumSolWtPar}

In this appendix we outline how we determine the solution parameters $\mu_{V}$ and $f_{V,V}$, which are used for the determination of the rest frame length density \eqref{eq:slenres} and the mean square velocity \eqref{eq:vLagWt} of the strings.
They can be estimated from the potential-weighted total energy $E$ \eqref{e:TotEne}, the potential-weighted potential energy $E_V$ \eqref{eq:EStrV}, the universe-frame string length $\ell_\text{w}$ \eqref{e:ellWinDef}, and the scalar velocity estimator $\vs$ \eqref{e:ellWinDef} by the inverse relations 
\bea
\mu_V &=&  - \frac{L}{\ell_\text{w}(1 - \vs^2)^{1/2}},\\
f_{V,V} &=& \frac{E_V}{E}.
\eea
These are 
applied to the string configurations at the end of the diffusive evolution, where the RMS velocity is very small.  The configurations are generated by a set of simulations (see Section \ref{sec:simulations}) with lattice sizes 1k, 2k, 4k and 8k, lattice spacings $\De x\in \{ 0.5, 0.25, 0.125, 0.0625 \}$, and initial field correlation length $\IniCorLen = 40$. For these simulations, the comoving string core width is 
$\ws^c = 0.5$, and the 
mean string separation at the end of diffusion is approximately  $35\ws^{c}$ (by both the universe-frame and rest-frame measures). We find the weighted string energy per unit length and potential energy fractions given in Table \ref{t:NumSolWtPar}. 
From these we determine functions $\mu_V( \Delta x/\ws^c)$ and $f_{V,V}(\Delta x /\ws^c)$ by least squares fits to a quadratic polynomial.

\begin{table}[hb]

\begin{tabular}{c c c}
$\Delta x/\ws^c$ & $\mu_V/\eta^2$ & $f_V$ \\
\hline
$1$ & $0.8517$ & $0.3732$ \\ 
$0.5$ & $0.8778$ & $0.3696$ \\ 
$0.25$ & $0.8896$ & $0.3687$ \\ 
$0.125$ & $0.8882$ & $0.3685$ 
\end{tabular}

\caption{\label{t:NumSolWtPar}
Potential-weighted string mass per unit length $\mu_V$ and fraction of potential-weighted potential energy in near-static axion string solutions, as a function of lattice spacing divided by comoving string width.
}

\end{table}

\section{Convergence for all simulation sizes}
\label{s:ConAll}

Here we give estimates of the asymptotic values for the string length density parameters $\zew$ (universe frame) and $\zer$ (string rest frame), and the RMS velocity $v_L$, for all simulations (box sizes 4k, 8k, 12k, and 16k).

Round symbols show estimates for the length density obtained by linear fits of the mean string separation $\xi(t)$ to cosmic time $t$.
Solid squares show the value at conformal time $\tau = L/2$, where $L$ is the simulation box side length in comoving coordinates. The corresponding cosmic time $t_{L/2}$ is given in the text boxes.
The empty squares mark values at cosmic time $t = t_{L/2}/2$. 

On the $x$ axis is plotted the value of  $\zer$ at the start of the physical evolution $t_\text{cg}$, averaged over all simulations at the selected box size and initial field correlation length, given in the legend. 

The olive green dashed line shows the mean of runs with $5 \le \IniCorLen \le 80$ for $\zew$ and $\zer$, which are held to have converged, and of all runs for $v_L$.  
The olive green band shows the 1$\si$ uncertainties (see Section \ref{sec:asymLv}). 

The grey dashed lines and bands show results from \cite{Hindmarsh:2019csc}, while the purple dashed lines and bands is the equivalent result from \cite{Hindmarsh:2021vih}.  

 \begin{figure}[h]
    \centering
     \includegraphics[width=\columnwidth]{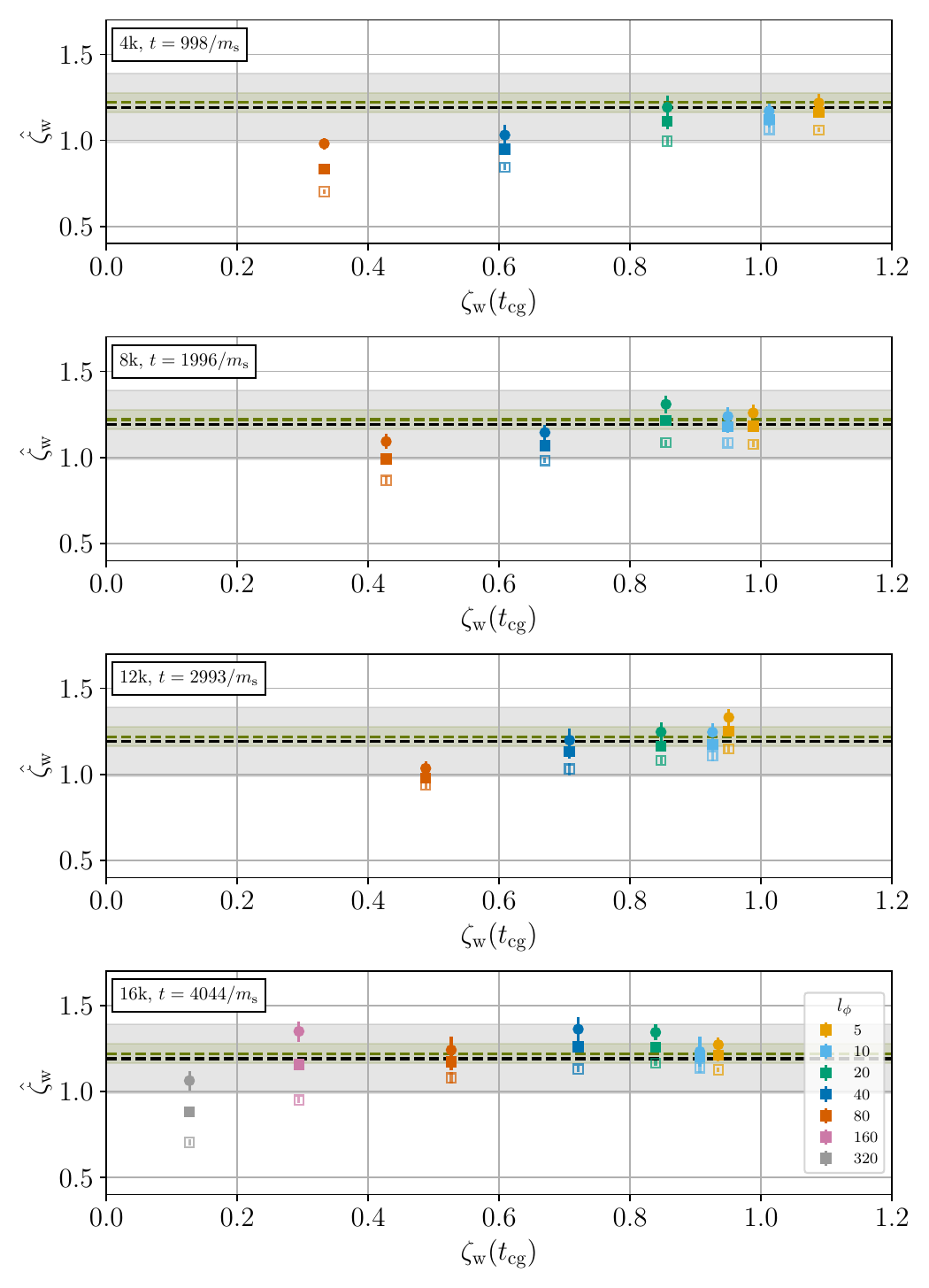}

    \caption{\label{fig:zeta_r_final_all} Estimates of asymptotic values of scaled universe frame length density parameter $\zew$, in all simulations (see text for an explanation of the symbols). The olive green dashed line and band shows the mean and 1$\si$ uncertainty of runs with $5 \le \IniCorLen \le 80$.  The grey dashed lines and bands is the equivalent result from \cite{Hindmarsh:2019csc}.  
}
 \end{figure}

 \begin{figure}[ht]
    \centering
     \includegraphics[width=\columnwidth]{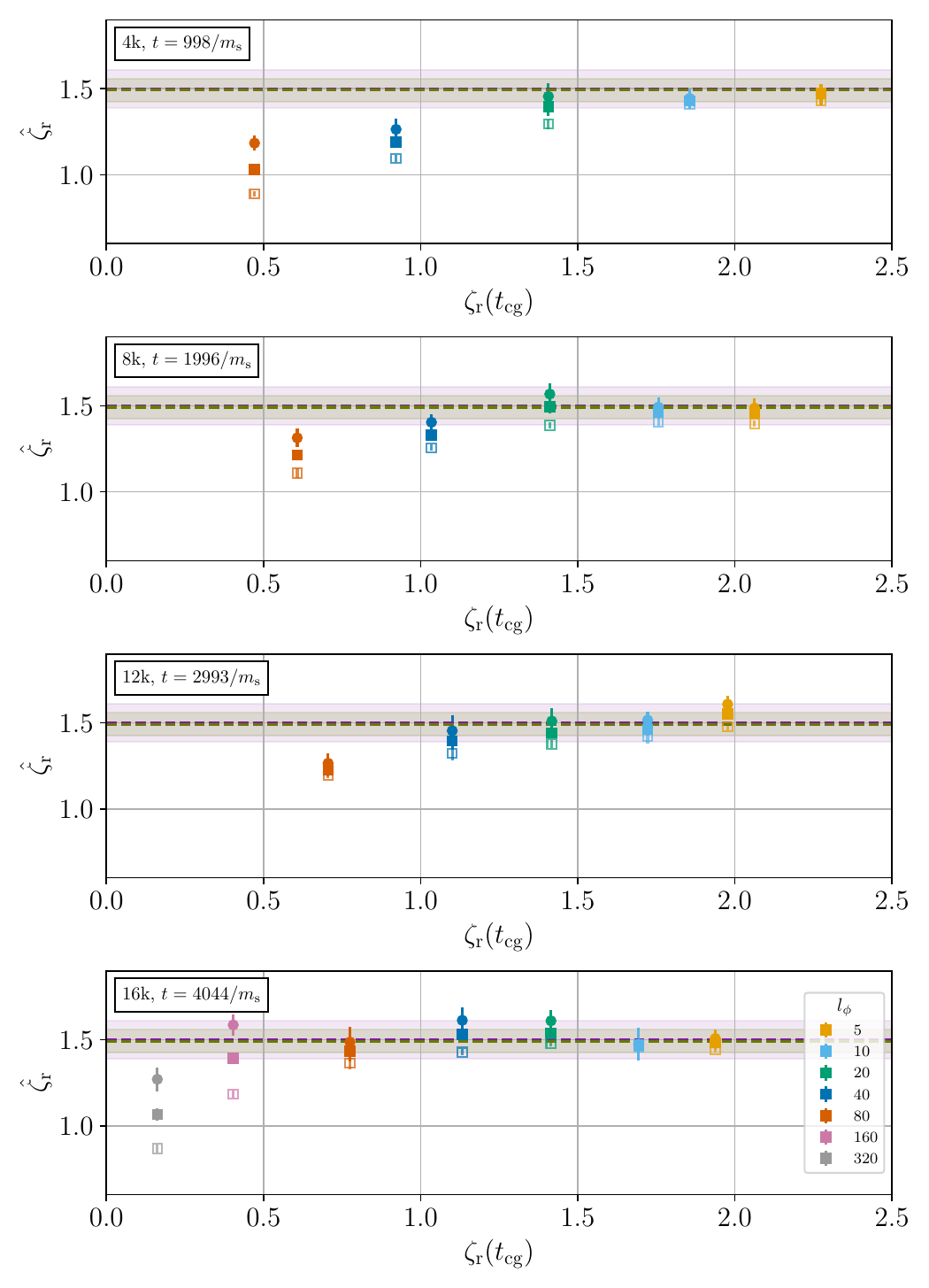}

    \caption{\label{fig:zeta_w_final_all} Estimates of asymptotic values of scaled rest frame length density parameter $\zer$, in all simulations (see text for an explanation of the symbols). The olive green dashed line and band shows the mean and 1$\si$ uncertainty of runs with $5 \le \IniCorLen \le 80$.
The purple dashed lines and bands is the equivalent result from \cite{Hindmarsh:2021vih}.  
}
 \end{figure}

 \begin{figure}[h]
    \centering
     \includegraphics[width=\columnwidth]{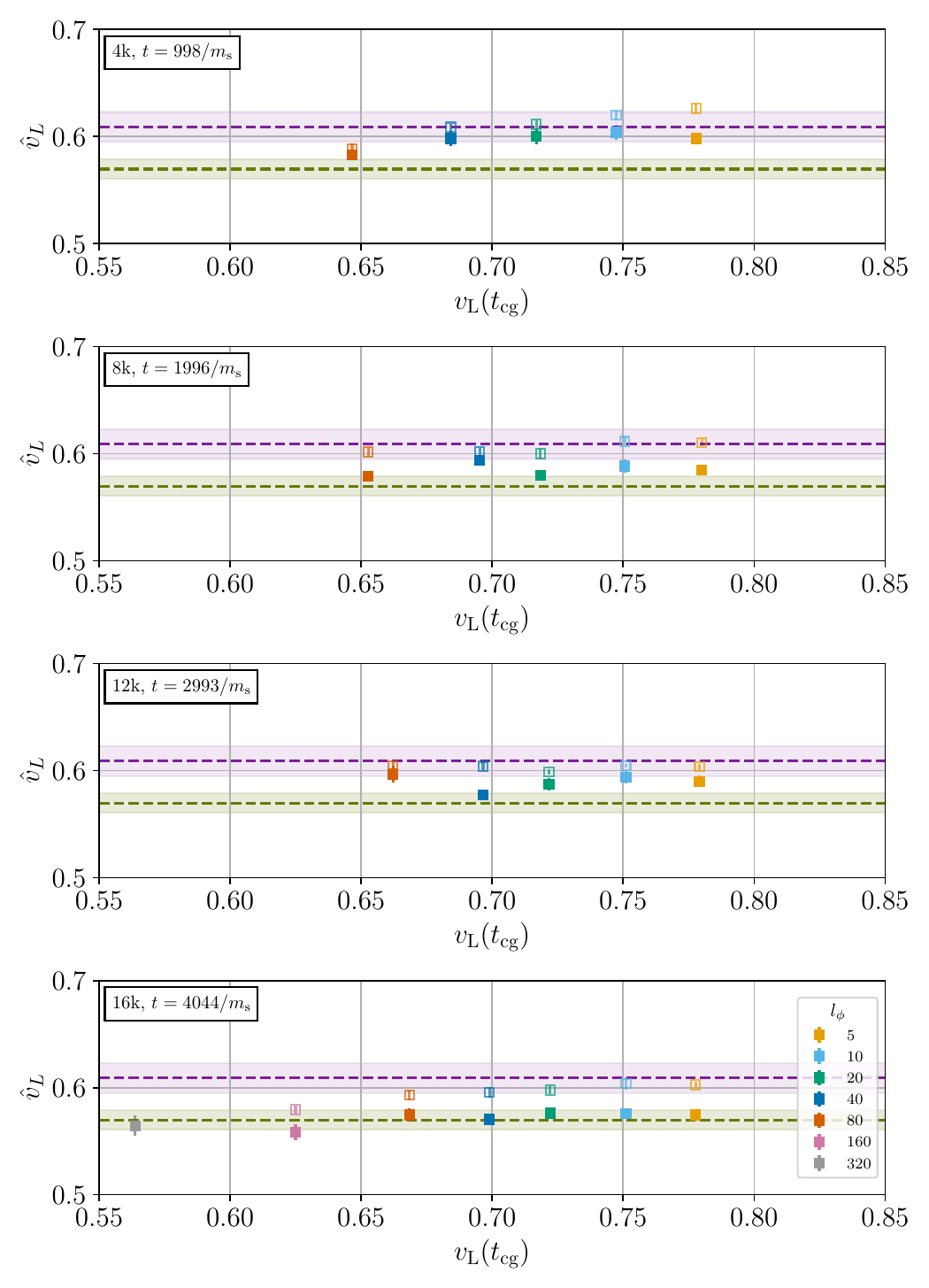}

    \caption{\label{fig:v_L_final_all} Estimates of asymptotic values of the RMS velocity $v_L$, in all simulations (see text for an explanation of the symbols). The olive green dashed line and band shows the mean and 1$\si$ uncertainty of runs with $5 \le \IniCorLen \le 320$. The purple dashed lines and bands is the equivalent result from \cite{Hindmarsh:2021vih}.  
}
 \end{figure}

\bibliography{axion} 

\end{document}